\documentclass[11pt]{article}

\usepackage{packages}
\usepackage{dsfont}
\usepackage{mathrsfs}
\usepackage{amsmath}
\usepackage{amsfonts}
\usepackage{amssymb}
\usepackage{graphicx}
\usepackage[colorinlistoftodos]{todonotes}

\begin{document}
\title{\bf On the equivalence theorem for integrable systems}

\author{ {\bf A. Melikyan}$\,^{1}$, {\bf E. Pereira}$\,^{2}$, {\bf V. O. Rivelles}$\,^{2}$ \thanks{\tt amelik@gmail.com, lilausp@gmail.com, rivelles@if.usp.br}\\

$^1$ \sl{Instituto de F\'{\i}sica}\\
\sl{Universidade de Bras\'{\i}lia, 70910-900, Bras\'{\i}lia, DF, Brasil}\\
\sl{and}\\
\sl{{International Center of Condensed Matter Physics} }\\
\sl{C.P. 04667, Brasilia, DF, Brazil} \\

$^2$ \sl{Instituto de F\'{\i}sica}\\
\sl{Universidade de S\~{a}o Paulo, C. Postal 66318, 05314-970, S\~{a}o Paulo, SP, Brasil\\}}

\date{\today}
  
\maketitle

\begin{abstract}
	We investigate the equivalence theorem for integrable systems using two formulations of the Alday-Arutyunov-Frolov model. We show that the S-matrix is invariant under the field transformation which reduces the non-linear Dirac brackets of one formulation into the standard commutation relations in the second formulation. We also explain how to perform the direct diagonalization of the transformed Hamiltonian by constructing the states corresponding to self-adjoint extensions.
\end{abstract}

\newpage
\newpage
\section{Introduction}\label{sec:intro}

The equivalence theorem \cite{Chisholm:1961,Kamefuchi:1961sb,Salam:1971sp} is a statement about the invariance of the S-matrix under field redefinitions in a quantum field theory. In this note we consider the equivalence theorem for the Alday-Arutyunov-Frolov ($AAF$) model \cite{Alday:2005jm,Arutyunov:2005hd}, which is an interesting fermionic integrable model arising from the  $\mathfrak{su}(1|1)$ subsector of string theory on $AdS_{5} \times S^{5}$ background, for which the full understanding of the classical and quantum integrability is still an open problem. Some of its classical and quantum integrability properties, such as the $S$-matrix factorization property, and classical integrability based on the inverse scattering method were considered in \cite{Klose:2006dd,Melikyan:2008ab,Melikyan:2012kj,Melikyan:2014yma,Melikyan:2014temp}. 

There are two formulations of the $AAF$ model, which were given in the original paper \cite{Alday:2005jm,Arutyunov:2005hd}. In the first formulation the Lagrangian is a two-dimensional purely fermionic quantum field theory, invariant under the Lorentz transformation, with a very involved structure of the Dirac brackets between the components of the fermionic field. This causes severe technical complications when investigating the integrability properties of the model. From the perturbative analysis point of view, the information contained in the Dirac brackets is not used, and while it is possible to show the  $S$-matrix factorization at the first loop order, there are serious computational barriers to generalize the analysis to higher loops. One would also like to obtain a non-perturbative proof of the $S$-matrix factorization \cite{Zamolodchikov:1978xm,Parke:1980ki,Dorey:1996gd}. From the point of view of the inverse scattering  method \cite{Faddeev:1988qp,Novikov:1984id,Korepin:1997bk,Essler:2005bk} the complicated structure of the Dirac brackets makes it hard to analyze the algebra of the Lax operator. In \cite{Melikyan:2012kj
,Melikyan:2014yma} it was shown that the classical algebra of Lax operators displays a non-ultralocal behavior \cite{Maillet:1985ek,Maillet:1985ec,Maillet:1985fn,Freidel:1991jv,Freidel:1991jx,Delduc:2012vq,Delduc:2012qb,Dorey:2006mx,Benichou:2010ts,Benichou:2011ch,Benichou:2012hc,Kundu:2003cu}.\footnote{The non-ultralocal structure of the Lax operator algebra for the $AAF$ model is such that it contains not only the first derivative of the delta-function, but also the second derivative. This leads to a more involved description of the algebra in terms of $(r,s_{1},s_{2})$ matrices. Nevertheless, it was also shown in \cite{Melikyan:2012kj,Melikyan:2014yma} that this complication leads only to a shift in $(r,s)$-pair, defining the classical algebra between the transition matrices.} While for simpler models classical integrability can be analyzed in detail, and the action-angle variables can be found \cite{Maillet:1985ek,Maillet:1985ec,Maillet:1985fn,Freidel:1991jv,Freidel:1991jx,Melikyan:2014yma}, for the $AAF$ model the non-linear structure of the Dirac bracket is too complicated, and the corresponding action-angle variables are still unknown. The quantization of such non-ultralocal models is still an unsolved problem \cite{Freidel:1991jv,Freidel:1991jx}.
 
In contrast, in the second formulation of the $AAF$ model \cite{Alday:2005jm}, one performs a fields transformation in order to reduce the complicated non-linear structure of the Dirac brackets to the standard canonical relations between the fermion field components. This is an obvious advantage for further investigation of the integrable properties of the $AAF$ model. It is not hard to obtain the corresponding Lax pair from the representation given in \cite{Melikyan:2012kj}. The corresponding algebraic structures for the Lax operator and the transition matrices are then expected to have a simpler form. The price, however, is the loss of manifest relativistic invariance in the resulting Lagrangian, which describes a very complicated interacting theory with up to the sixth order terms in the fermion fields and their derivatives. It is not at all obvious \emph{a priori} that the $S$-matrices of both theories, related by a transformation of the fields, will be the same, thus, guaranteeing the quantum integrability of the transformed Lagrangian.\footnote{This is especially surprising since the original theory is a relativistic theory, while the transformed theory is not.} The demonstration of this fact is the essence of the equivalence theorem. 

More importantly, a general proof of the equivalence theorem would allow us to consider a transformation of the fields, possibly a non-local one, to reduce the original non-ultralocal theory to an ultralocal one, without affecting the quantum integrability of the theory. One already has the freedom to use a gauge transformation of the Lax pair in order to simplify the algebra of Lax operators (for details of this approach see \cite{Freidel:1991jv,Freidel:1991jx}). However, the equivalence theorem would guarantee the $S$-matrix factorization property allowing the consideration of a much larger class of field transformations in order to reduce a given non-ultralocal model to an ultralocal one.

There exists a number of different proofs and approaches to the equivalence theorem (see \cite{Lam:1972mb,Lam:1973qa,Lam:1973qk,Lam:1974dd,Rouet:1974qm,Sharatchandra:1978dj,Chicherin:2011ad,Ferrari:2002kz,Furnstahl:2000we,Tyutin:2000ht,Blasi:1999as,Kilian:1994mg,Arzt:1993gz} and the references therein), and, to the best of our knowledge, the most complete and simple proof for renormalizable theories has been given in \cite{Bergere:1975tr}. The proof assumes that the theory is renormalized within the $BPHZ$ scheme \cite{Itzykson:1980rh}, and is based on a perturbative analysis of the invariance of Green's functions under field transformations.\footnote{The equivalence theorem and the problems arising from the non-perturbative analysis has been considered in \cite{Sharatchandra:1978dj} where it has been shown that  in general the equivalence theorem does not hold.}

There is so far no strict proof of renormalizability of the $AAF$ model. It is certainly not power-counting renormalizable, due to derivatives present in the interaction vertices. The infinite number of symmetries imposed by quantum integrability are nevertheless expected to render the model to be renormalizable. However, the full proof of quantum integrability is also missing. The $S$-matrix factorization has been shown only up to one-loop order \cite{Melikyan:2011uf}, and the inverse scattering method cannot be directly generalized, although some progress has been made in this direction \cite{Melikyan:2012kj,Melikyan:2014yma}. 

In this paper we show that the quantum Hamiltonian of the $AAF$ model, for the second formulation, can be diagonalized by means of the coordinate Bethe ansatz in the \emph{pseudo-vacuum}, if one employs a suitable regularization of the Hamiltonian in order to avoid singular operator products at the same point.\footnote{It is much harder to use the coordinate Bethe ansatz for the first formulation. In this case it is necessary first to regularize the complex structure of the Dirac brackets.} This already gives a strong indication that the model is indeed renormalizable, although a more complete analysis should still be done in order to construct the physical vacuum, states and the $S$-matrix as in \cite{Korepin:1979qq,Korepin:1979hg}. In the process of diagonalization we derive the explicit wave-functions and reproduce the $S$-matrix for the original formulation of the $AAF$ model \cite{Klose:2006dd,Melikyan:2008ab}. We also trace the essential steps of the proof of the equivalence theorem and adapt it for our case. The key step of the proof is the required renormalization of the wave-functions in order to guarantee the invariance of the $S$-matrix. We then relate it with the regularization of the quantum Hamiltonian and the corresponding renormalization of the wave-functions in the quantum-mechanical description. Finally, we give a third interpretation of the results by showing that the diagonalization and the form of the wave-function correspond to self-adjointness of the quantum-mechanical Hamiltonian, which requires the construction of self-adjoint extensions similarly to the case considered in \cite{Melikyan:2008ab,Melikyan:2010fr} for the Landau-Lifshitz model.

Our paper is organized as follows. In section \ref{sec:aafov} we give a summary of the known results related to the integrability properties of the $AAF$ model. In section \ref{sec:eq} we briefly recap the proof of \cite{Bergere:1975tr}, adapting it to our case and emphasizing only the essential differences arising in this case. In section \ref{sec:diag} we explain how to diagonalize the Hamiltonian of the $AAF$ model in the second formulation, and show that it requires a regularization of the quantum Hamiltonian, as well as a subtle construction of eigenstates. In section \ref{sec:sa} we show that this construction corresponds to finding self-adjoint extensions of the Hamiltonian, and make a connection between the direct diagonalization and the perturbative proof via renormalization of the wave-functions. In the last section, we discuss some open problems and future directions. Finally, in the Appendix, we collect some explicit formulas and computational details used in the main text.

\section{Alday-Arutyunov-Frolov model: Overview}\label{sec:aafov}

In this section we give a brief overview of the $AAF$ model, and discuss its most essential properties. A more complete account is given in \cite{Alday:2005jm,Arutyunov:2005hd,Melikyan:2011uf,Melikyan:2012kj,Melikyan:2014yma}. The model arises from strings on $AdS_{5} \times S^{5}$ background by reduction to the $su(1|1)$ subsector, where the bosonic degrees of freedom are eliminated in favor of the fermionic ones. 

The Lagrangian of the $AAF$ model, which is a two-dimensional fermionic model, has the form:
\begin{align}
	\label{aafov:lagrangian} \mathscr{L}_{AAF}  = &-J-\frac{i}{2}(\bar{\psi}\rho^{0}\partial_{0}\psi-\partial_{0}\bar{\psi}\rho^{0}\psi)+i\frac{\sqrt{\lambda}}{2J}(\bar{\psi}\rho^{1}\partial_{1}\psi-\partial_{1}\bar{\psi}\rho^{1}\psi)+\bar{\psi}\psi\nonumber \\
 & +\frac{\sqrt{\lambda}g_{2}}{4J^{2}}\varepsilon^{\alpha\beta}(\bar{\psi}\partial_{\alpha}\psi\bar{\psi}\rho^{5}\partial_{\beta}\psi-\partial_{\alpha}\bar{\psi}\psi\partial_{\beta}\bar{\psi}\rho^{5}\psi)-\frac{\sqrt{\lambda}g_{3}}{16J^{3}}\epsilon^{\alpha\beta}(\bar{\psi}\psi)^{2}\partial_{\alpha}\bar{\psi}\rho^{5}\partial_{\beta}\psi.
\end{align}
Here, $\alpha, \beta = 0,1$, $\lambda$ is the t'Hooft coupling, $J$ is the total angular momentum of the string in $S^5$, and the following representation of the Dirac matrices is used:
\begin{equation}
\rho^{0}=\left(\begin{array}{cc}
-1 & 0\\
0 & 1
\end{array}\right),\quad\rho^{1}=\left(\begin{array}{cc}
0 & i\\
i & 0
\end{array}\right), \quad\rho^{5}=\rho^{0}\rho^{1}.\label{aafov:rho_matrcies}
\end{equation}
After some rescaling of the fields, the action can be written in the explicitly relativistic form (for details see \cite{Alday:2005jm,Melikyan:2011uf}):
\begin{align}
	 S &= \int dy^0 \: \int _0^J dy^1 \: \left[ i \bar{\psi} \delslash \psi \: - m \bar{\psi} \psi + \frac{g_2}{4m} \epsilon^{\alpha \beta} \left( \bar{\psi}
	\partial_{\alpha} \psi \; \bar{\psi}\: \gamma^3 
	\partial_{\beta} \psi -
	\partial_{\alpha}\bar{\psi} \psi \; 
	\partial_{\beta} \bar{\psi}\: \gamma^3 \psi \right) \right.- \nonumber \\
	&- \left. \frac{g_3}{16m} \epsilon^{\alpha \beta} \left(\bar{\psi}\psi\right)^2 
	\partial_{\alpha}\bar{\psi}\:\gamma^3
	\partial_{\beta}\psi \right], \label{aafov:lag_relativistic}
\end{align}
where the Dirac matrices $\gamma^{\mu}$ have the form:\footnote{The matrices $\rho^{\mu}$ and $\gamma^{\mu}$ in \eqref{aafov:rho_matrcies} are related by the following transformation: 
\begin{equation}
\gamma^{\mu} = M \rho^{\mu} M^{-1}, \quad M = \frac{1}{\sqrt{2}} \left( 
	\begin{array}{cc}
		1 & -i \\
		-1 & -i 
	\end{array}
	\right).
\end{equation}	
	}
\begin{equation}
	\label{aafov:gamma_matrices} \gamma^0 = \left( 
	\begin{array}{cc}
		0 & 1 \\
		1 & 0 
	\end{array}
	\right), \quad \gamma^1 = \left( 
	\begin{array}{cc}
		0 & -1 \\
		1 & 0 
	\end{array}
	\right), \quad \gamma^3 = \gamma^0 \gamma^1.
\end{equation}

The coupling constants $g_{2}$ and $g_{3}$ were introduced in \cite{Melikyan:2011uf}, where it was shown that the quantum integrability, i.e., the $S$-matrix factorization property up to the first loop approximation holds, provided the relation $g_{2}^{2} = g_{3}$ between the coupling constants is satisfied. The same condition was also shown in \cite{Melikyan:2012kj} to guarantee  the classical integrability of the model. 

The two-particle scattering $S$-matrix has the form:
 \begin{equation}
S(\theta_{1},\theta_{2})=\frac{1-\frac{img_{2}}{4}\sinh(\theta_{1}-\theta_{2})}{1+\frac{img_{2}}{4}\sinh(\theta_{1}-\theta_{2})},\label{aafov:smatrix}
 \end{equation}
where $\theta_{1}$ and $\theta_{2}$ are the rapidities of the scattered particles with the momenta $p^{1}=m\sinh{\theta_{1}}$ and $p^{2}=m\sinh{\theta_{2}}$. The dependence of the $S$-matrix on the difference of the rapidities is a consequence of the relativistic invariance of the Lagrangian \eqref{aafov:lag_relativistic}. Moreover, the above $S$-matrix \eqref{aafov:smatrix} has been obtained in the pseudo-vacuum, defined by the equation:
\begin{equation}
	\psi(x) |0 \rangle=0. \label{aafov:vacuum}
\end{equation}
The physical vacuum can be reconstructed using the standard methods \cite{Korepin:1997bk,Korepin:1979qq,Korepin:1979hg}. In the next sections we will discuss this point and the implications of this choice in more details.

The structure of the Dirac brackets for the components of the fermionic field  $\psi$ has a very complicated form (see \cite{Alday:2005jm,Melikyan:2012kj} for details\footnote{Notice that there were some missing factors in the Lagrangian given in \cite{Alday:2005jm}, which propagated into the Dirac structure as well. It was corrected later in \cite{Melikyan:2012kj}. We do not write here the lengthy formulas of Dirac brackets, and instead refer the reader to the appendix {\bf F} of \cite{Melikyan:2012kj} for the explicit expressions.}), extending to the sixth order in the fermion field components and their derivatives. The  non-trivial structure of the Dirac brackets makes the analysis of the $AAF$ model a difficult problem in two aspects. Firstly, the perturbative calculations seem to be impossible to continue beyond one loop, due to extremely involved diagrammatic calculations and cancelations mechanism, responsible for the factorization of the $n$-particle scattering $S$-matrix. Most importantly, the attempts to develop the inverse scattering method for the $AAF$ model leads to serious barriers when trying to obtain, for instance, the lattice version of the model. Indeed, as shown in \cite{Melikyan:2012kj,Melikyan:2014yma} the $AAF$ model is a non-ultralocal model (see  \cite{Maillet:1985ek,Maillet:1985ec,Maillet:1985fn,Freidel:1991jv,Freidel:1991jx,Kundu:2003cu} and the references therein). In fact the algebra of the Lax pair has a more complex structure and contains terms proportional not only to the first derivative of the delta-function, but also to the second derivative of the delta-function. In the former case, the non-ultralocal algebra can be described using a pair of matrices - the $(r,s)$ pair which was considered in details in \cite{Maillet:1985ek}. In the case of integrable models which contain the second derivatives of the delta-function, one has also to introduce, as was shown in \cite{Melikyan:2014yma}, the third matrix in order to completely describe the algebra of the Lax operator. For the $AAF$ model the exact form of the matrices  $(r,s_{1}, s_{2})$ has a very complicated non-linear character \cite{Melikyan:2012kj}, which is the direct consequence of the involved Dirac bracket structure of the theory. Moreover, one is interested in the algebra of the transition matrices, which can be found via a symmetrization procedure proposed by Mailett in \cite{Maillet:1985ek}, and is expressed in terms of the $(u,v)$ pair, which in turn can be found from the $(r,s)$ pair. Remarkably, in the more complex case described by  the $(r,s_{1}, s_{2})$ matrices, the algebra of transition matrices has the same form as in  the simpler case described by the $(r,s)$ pair, and can also be expressed in terms of only two matrices - the $(u',v')$ pair \cite{Melikyan:2014yma}. One can then construct the angle-action variables following the standard procedure \cite{Faddeev:1988qp,Novikov:1984id,Korepin:1997bk,Essler:2005bk}. This program has been realized for simpler models in \cite{Melikyan:2014yma}. For the $AAF$ model one faces serious technical difficulties and, as we mentioned above, this is largely due to the non-linear structure of the Dirac brackets. 

An alternative approach is to reduce the non-linear Dirac brackets to the standard canonical Poisson relations by means of some field redefinitions before calculating the Lax algebra.  As shown in \cite{Alday:2005jm}, such a reduction is indeed possible, and the corresponding transformation has the form:
\begin{align}
\psi & \rightarrow\psi
+\frac{a_{1}}{J^{2}}\rho^{1}\psi(\partial_{1}\bar{\psi}\psi)
+\frac{a_{2}}{J^{3}}\rho^{1}\partial_{1}\psi(\bar{\psi}\psi)^{2}
+\frac{a_{3}}{J^{4}}\partial_{1}^{2}\psi(\bar{\psi}\psi)^{2}
+\frac{a_{4}}{J^{4}}\partial_{1}\psi(\partial_{1}\bar{\psi}\psi-\bar{\psi}\partial_{1}\psi)\bar{\psi}\psi\nonumber \\
 &
+\frac{a_{5}}{J^{4}}\rho^{0}\psi(\partial_{1}\bar{\psi}\rho^{0}\partial_{1}\psi)\bar{\psi}\psi+\frac{a_{6}}{J^{6}}\rho^{1}\partial_{1}\psi(\partial_{1}\bar{\psi}\partial_{1}\psi)(\bar{\psi}\psi)^{2},\nonumber
\\
\bar{\psi} & \rightarrow\bar{\psi}
+\frac{a_{1}^{*}}{J^{2}}\bar{\psi}\rho^{1}(\bar{\psi}\partial_{1}\psi)
+\frac{a_{2}^{*}}{J^{3}}\partial_{1}\bar{\psi}\rho^{1}(\bar{\psi}\psi)^{2}
+\frac{a_{3}}{J^{4}}\partial_{1}^{2}\bar{\psi}(\bar{\psi}\psi)^{2} +\frac{a_{4}}{J^{4}}\partial_{1}\bar{\psi}(\bar{\psi}\partial_{1}\psi-\partial_{1}\bar{\psi}\psi)\bar{\psi}\psi \nonumber \\
 &
+\frac{a_{5}}{J^{4}}\bar{\psi}\rho^{0}(\partial_{1}\bar{\psi}\rho^{0}\partial_{1}\psi)\bar{\psi}\psi+\frac{a_{6}^{*}}{J^{6}}\partial_{1}\bar{\psi}\rho^{1}(\partial_{1}\bar{\psi}\partial_{1}\psi)(\bar{\psi}\psi)^{2}, \label{aafov:trans_psi}
\end{align}
where the coefficients $a_{1} \ldots a_{6}$ are:
\begin{align}
a_{1}=&i\frac{\sqrt{\lambda}}{2} g_{2},\ a_{2}=-i\frac{\sqrt{\lambda}}{16} g_{3},\ a_{3}=-\frac{\lambda}{16}g_{2}^{2},\nonumber
\\ a_{4}=&-\frac{{\lambda}}{8}g_{2}^{2},\ a_{5}=\frac{{\lambda}}{8}g_{2}^{2},\ a_{6}=-i\frac{5 \, {\lambda}^{3/2}}{32}g_{2}^{3},\label{aafov:a_coefficients}
\end{align}
and $a_{i}^{*}$ denote the complex conjugate of $a_{i}$. We also note that there exists the inverse to \eqref{aafov:trans_psi} transformation.\footnote{ Its exact form can be found in \cite{Alday:2005jm}.} We will use this fact when we discuss the equivalence theorem in the next section.

Using the field redefinitions \eqref{aafov:trans_psi} and \eqref{aafov:a_coefficients}, one can show that the transformed Lagrangian takes the form:
\begin{align}\label{aafov:lagrang_transformed}
\mathscr{L}  &=  i(\bar{\psi}{\gamma}^{\alpha}\partial_{\alpha}\psi)
-m\bar{\psi}\psi
+\frac{i}{2}g_{2}(\bar{\psi}\psi)(\bar{\psi}\gamma^{1}\partial_{1}\psi-\partial_{1}\bar{\psi}\gamma^{1}\psi)+\frac{g_{2}}{2m}\left[(\bar{\psi}\partial_{1}\psi)^{2}+(\partial_{1}\bar{\psi}\psi)^{2}\right]  \\
&-\frac{g_{3}}{8m}(\bar{\psi}\psi)^{2}(\partial_{1}\bar{\psi}\partial_{1}\psi)-\frac{g_{2}^{2}}{4m}(\bar{\psi}\psi)^{2}(\partial_{1}\bar{\psi}\partial_{1}\psi)-i\frac{g_{2}^{2}}{8m^{2}}(\bar{\psi}\psi)^{2}(\partial_{1}\bar{\psi}\gamma^{1}\partial_{1}^{2}\psi-\partial_{1}^{2}\bar{\psi}\gamma^{1}\partial_{1}\psi)\nonumber \\
&-i\frac{g_{2}^{2}}{2m^{2}}(\bar{\psi}\psi)(\bar{\psi}\partial_{1}\psi-\partial_{1}\bar{\psi}\psi)(\partial_{1}\bar{\psi}\gamma^{1}\partial_{1}\psi)-\frac{g_{2}^{3}}{2m^{3}}(\bar{\psi}\psi)^{2}(\partial_{1}\bar{\psi}\partial_{1}\psi)^{2.}\nonumber
\end{align}

The new Hamiltonian can be easily read off from this expression (see \eqref{opreg:hamiltonian} below), and its complicated form is the trade off for the standard canonical Poisson brackets for the fermion fields. The transformation \eqref{aafov:trans_psi} can be also used to obtain the Lax pair, from the Lax pair given in \cite{Melikyan:2012kj} for the transformed Lagrangian \eqref{aafov:lagrang_transformed}. It is expected that due to the simpler canonical Poisson brackets of the transformed theory \eqref{aafov:lagrang_transformed}, the algebra of Lax operators will  have a simpler form. It is, however, not \emph{a priori} obvious, that the quantum theories corresponding to the original \eqref{aafov:lagrangian} and the transformed \eqref{aafov:lagrang_transformed} Lagrangians are equivalent. Here we should specify what we mean by 
\emph{equivalent}. The key object that encodes the integrability of the model, i.e., the spectrum, is the $n$-particle scattering $S$-matrix. For the original theory, the $2$-particle scattering $S$-matrix is given in \eqref{aafov:smatrix}. If the model is integrable for the quantum theory (this has been checked in \cite{Melikyan:2011uf} for the Lagrangian \eqref{aafov:lagrangian} in the one-loop approximation), then the $n$-particle $S$-matrix can be found from \eqref{aafov:smatrix} and the $S$-matrix factorization property. There is no strict proof, however, of the quantum integrability, and the most reliable method, the inverse scattering method, has not yet been generalized for the quantum case for non-ultralocal theories. One hopes to simplify this non-ultralocal algebra for the theory \eqref{aafov:lagrang_transformed} written in terms of the transformed fields \eqref{aafov:trans_psi}. It is not, however, immediately obvious that this classical transformation will guarantee the invariance of the $S$-matrix in the quantum theory. This is the general statement of the equivalence theorem, which we will discuss for our case in the next section. In section \ref{sec:diag} we will also confirm our results by performing an explicit diagonalization of the quantum Hamiltonian corresponding to the transformed theory \eqref{aafov:lagrang_transformed}. 

\section{The equivalence theorem: $S$-matrix invariance}\label{sec:eq}

In this section we follow the proof of \cite{Bergere:1975tr}, modifying it appropriately for our case, to show the invariance of the $S$-matrix under the field transformations \eqref{aafov:trans_psi}. As we mentioned above, due to the Lorentz invariance of the original theory \eqref{aafov:lag_relativistic}, the $n$-particle scattering $S$-matrix of the original theory depends only on the difference of the rapidities of the scattered particles (see \eqref{aafov:smatrix}). Hence, it is not evident that the $S$-matrix will retain this form under the field transformations \eqref{aafov:trans_psi},  resulting in the Lagrangian \eqref{aafov:lagrang_transformed} which does not have an explicit relativistic invariant form. 

The proof of the theorem for a fermionic model is similar to the case of a single bosonic field given in \cite{Bergere:1975tr}. There are, however, some differences, and therefore we briefly retrace the steps of the proof adapting it for the fermionic case. We also give a simpler derivation of one of the key relations (see \eqref{smatrix:pole_i_alt} below), which leads to the invariance of the $S$-matrix. We stress here that the following proof is valid only for renormalizable theories, and in the case of the $AAF$ model the proof of renormalizability is so far missing. In fact, it is not even a power-counting renormalizable theory.  It is, however, supposed to be renormalizable due to integrability and the infinite symmetries associated with it \cite{Alday:2005jm,Klose:2006dd}. In the next section we will show that the quantum Hamiltonian can be diagonalized via the coordinate Bethe ansatz, suggesting renormalizability of the theory. 

We start by showing that the Green functions of the original and transformed Lagrangians are invariant under the field redefinitions.  Using this result, one shows that the $S$-matrix of the two Lagrangians is also invariant, up to some renormalization for the fields. We first represent the Lagrangian \eqref{aafov:lag_relativistic} in a more general form:
\begin{align}
\mathscr{L}(\psi, \bar{\psi}) & =i\bar{\psi}\gamma^{\alpha}\partial_{\alpha}\psi-m\bar{\psi}\psi+\lambda\mathscr{L}_{int}\label{green:lagrang_generic}
\end{align}
where $\mathscr{L}_{int}$ is the interaction term, the Dirac matrices are given by \eqref{aafov:gamma_matrices}, and $\lambda$ is a coupling constant. One would like to show that the $S$-matrix is invariant under the change of variables:
\begin{align}
\psi\rightarrow\psi+F(\psi,\bar{\psi}),\label{green:t1} \\
\bar{\psi}\rightarrow\bar{\psi}+\bar{F}(\psi,\bar{\psi}),\label{green:t2}
\end{align}
where $\bar{F}(\psi,\bar{\psi})=F^{\dagger}(\psi,\bar{\psi})\gamma^{0}$. The function $F(\psi,\bar{\psi})$, which can be easily read off from the formulas \eqref{aafov:trans_psi} and \eqref{aafov:a_coefficients}, is a polynomial in $\psi$, $\bar{\psi}$ and the its derivatives, and satisfies the following restriction $F(\psi,\bar{\psi})\neq-\psi$. We will denote the transformed Lagrangian by $\mathscr{L}_{T}(\psi,\bar{\psi})$,
i.e.:
\begin{equation}
\mathscr{L}_{T}(\psi,\bar{\psi})=\mathscr{L}\left(\psi+F(\psi,\bar{\psi}),\bar{\psi}+\bar{F}(\psi,\bar{\psi})\right). \label{green:lagrang_T}
\end{equation}
To show that the Green function is invariant under the change of
variables \eqref{green:t1} and \eqref{green:t2}, one considers the transformations:
\begin{align}
\psi^{(\rho)}(\psi,\bar{\psi})=& \psi+\rho F(\psi,\bar{\psi}),\label{green:t1-1} \\
\bar{\psi}^{(\rho)}(\psi,\bar{\psi})=& \bar{\psi}+\rho\bar{F}(\psi,\bar{\psi}),\label{green:t2-1}
\end{align}
where the parameter $\rho \in [0,1]$. For such a transformation, we have the corresponding Lagrangian:
\begin{equation}
\mathscr{L}_{\rho}(\psi,\bar{\psi})=\mathscr{L}\left(\psi^{(\rho)}(\psi,\bar{\psi}),\bar{\psi}^{(\rho)}(\psi,\bar{\psi})\right). \label{green:lagrang_rho}
\end{equation}
For $\rho=0$ one obtains the original Lagrangian (\ref{green:lagrang_generic}) and for $\rho=1$ one finds the transformed Lagrangian (\ref{green:lagrang_T}).

The main result of \cite{Bergere:1975tr} is that the Green functions are invariant under the field redefinition that relates the original and transformed theories. In appendix \ref{sec:appendix} we give the derivation of this result adapted to our case of a fermion field. Explicitly, the relation has the form:
\begin{equation}
	\left\langle T\left[\prod_{i=1}^{s}\left\{ \zeta_{i}(\psi,\bar{\psi})\bar{\zeta}_{i}(\psi,\bar{\psi})\right\} (x_{i})\right]\right\rangle ^{\mathscr{L}}=\left\langle T\left[\prod_{i=1}^{s}\left\{ \zeta_{i}(\psi+F,\bar{\psi}+\bar{F})\bar{\zeta}_{i}(\psi+F,\bar{\psi}+\bar{F})\right\} (x_{i})\right]\right\rangle _{M}^{\mathscr{L}_{T}},\label{green:green}
\end{equation}
where $\zeta_{1}(\psi,\bar{\psi}),\zeta_{2}(\psi,\bar{\psi}),\ldots, \zeta_{s}(\psi,\bar{\psi})$ are arbitrary functions of fields. The indices $\mathscr{L}$ and $\mathscr{L}_{T}$ in \eqref{green:green} indicate that the correlation functions in \eqref{green:green} are  calculated with respect to the Lagrangians $\mathscr{L}$ \eqref{green:lagrang_generic} and $\mathscr{L}_{T}$ \eqref{green:lagrang_T} respectively. The index $M$ in the right-hand side of \eqref{green:green} is the order of the expansion in the parameter $\rho$. The $S$-matrix invariance under the field redefinitions follows from this relation, up to some field renormalization. This field renormalization will be reinterpreted in the next section by means of the direct diagonalization of the quantum Hamiltonian $\mathcal{H}_{\Delta}$ (see \eqref{opreg:hamiltonian} below). 

First, we recall that all the perturbative calculations in \cite{Klose:2006dd,Melikyan:2011uf} for the $AAF$ model with the original Lagrangian \eqref{aafov:lag_relativistic} were performed with respect to the \emph{false} vacuum \eqref{aafov:vacuum}. As we commented earlier, the $S$-matrix \eqref{aafov:smatrix} has been obtained exactly in this pseudo-vacuum, from which one can in principle obtain the physical $S$-matrix by the procedure outlined in \cite{Korepin:1979qq,Korepin:1979hg}. We will show below that for the transformed Lagrangian \eqref{aafov:lagrang_transformed}, and generally for the family $\mathscr{L}_{\rho}$ corresponding to the field transformations \eqref{green:t1-1} and \eqref{green:t2-1}, the vacuum can also be chosen to be the pseudo-vacuum defined as in \eqref{aafov:vacuum}.

The choice of the false vacuum is motivated by the fact that the perturbative calculations are significantly simpler to carry out. This is due to the structure of the free propagator described by the Lagrangian of the original theory $\mathscr{L}(\psi,\bar{\psi})$  \eqref{aafov:lag_relativistic}, which in the false vacuum \eqref{aafov:vacuum} is a purely retarded propagator $D_{ret}(x-y)$. It has the following form:
\begin{align}
	\label{smatrix:propagator_free} D_{ret}(x-y) &= \langle \Omega^{(0)}| T \psi(x) \bar{\psi}(y)| \Omega^{(0)}\rangle_{free} =\left(i \delslash +m \right) \int \frac{d^2p}{4 \pi^2}\: \frac{i e^{-i p\cdot (x -y)}}{p^2 - m^2+2i\varepsilon p_0},
\end{align}
where we have denoted the pseudo-vacuum of the original theory by $\Omega^{(0)}$, satisfying the condition:
\begin{equation}
	\psi(x)|\Omega^{(0)}\rangle =0. \label{smatrx:pseudo_vacuum_original_theory}
\end{equation}
A few comments are in order. First, the pole prescription is still invariant under Lorentz transformations. More importantly, as explained in \cite{Klose:2006dd}, one can readily prove a non-renormalizability theorem, which states that the ground state energy is not renormalized, the one-point Green function does not acquire  any quantum correction, and the two-point scattering matrix is given by the sum of bubble diagrams. The latter calculation is usually quite simple to carry out,\footnote{See \cite{Das:2007tb,Melikyan:2008ab,Melikyan:2011uf} for a general method to calculate such bubble diagrams. The key point is that in the false vacuum the one-loop correction to the two-particle scattering is proportional to the tree vertex, where two external lines can be taken off-shell. This allows one to easily calculate the scattering process for bubble diagrams with an infinite number of loops.} and, assuming the quantum integrability of the theory, it is enough to obtain the $n$-particle scattering matrix. 

For the proof of the equivalence theorem, it is the non-renormalizability of the one-point Green function that plays a key role. Indeed, as explained in \cite{Bergere:1975tr}, one must assume that the original Lagrangian is such that the full two-point function has a simple pole at $\dslash{p}=m$. In our case, due to the false vacuum and the non-renormalization theorem of the one-point Green function, this assumption is not needed - the full propagator coincides with the free propagator and has the form \eqref{smatrix:propagator_free}. Then, choosing the functions $\zeta_{1}$ and $\zeta_{2}$ in \eqref{green:trans} as follows:
\begin{align}
	\zeta_{1}\left(\psi^{(\rho)},\bar{\psi}^{(\rho)}\right)(x)&=\psi^{(\rho)}(\psi,\bar{\psi})(x),\label{smatrix:zeta_1} \\
	\zeta_{2}\left(\psi^{(\rho)},\bar{\psi}^{(\rho)}\right)(y)&=\bar{\psi}^{(\rho)}(\psi,\bar{\psi})(y),\label{smatrix:zeta_2}
\end{align}
one obtains from the general formula \eqref{green:zero}:
\begin{equation}
	\langle \Omega^{(\rho)} | T \Big( \psi+\rho F(\psi,\bar{\psi}) \Big)(x) \Big( \bar{\psi}+\rho \bar{F}(\psi,\bar{\psi}) \Big)(y) | \Omega^{(\rho)} \rangle  =\left(i \delslash +m \right) \int \frac{d^2p}{4 \pi^2}\: \frac{i e^{-i p\cdot (x -y)}}{p^2 - m^2+2i\varepsilon p_0}.\label{smatrix:simple_pole}
\end{equation}
To simplify our notations we have denoted in the above formula: $\psi^{(\rho)} \equiv \psi^{(\rho)}(\psi,\bar{\psi})$, $\bar{\psi}^{(\rho)} \equiv \bar{\psi}^{(\rho)}(\psi,\bar{\psi})$, and have omitted here and in what follows the indices $M$ and ${\mathscr{L}_{\rho}}$. We have also explicitly denoted  by $\Omega^{(\rho)}$ the vacuum of the transformed theory corresponding to $\mathscr{L}_{\rho}$. We stress that the right-hand side of the equation \eqref{smatrix:simple_pole} is invariant under the orthochronous Lorentz transformations. Thus, despite the fact that the field transformations given in \eqref{green:t1-1}, \eqref{green:t2-1} and \eqref{aafov:trans_psi} are not Lorentz invariant,\footnote{This implies of course that the Lagrangian \eqref{aafov:lagrang_transformed} is also not Lorentz invariant.} the relation \eqref{smatrix:simple_pole} implies that the Green's function on the left-hand side is still Lorentz invariant. In the next section we will use this fact explicitly when diagonalizing the quantum Hamiltonian of the model and when choosing a solution of the quantum-mechanical wave-function consistent with this requirement. This will also explain the fact that the $S$-matrix for the transformed theory depends on the difference of the rapidities as in \eqref{aafov:smatrix}.

One can use the formula \eqref{smatrix:simple_pole}, and proceed as in \cite{Bergere:1975tr} to derive the following key relation:
\begin{align}\label{smatrix:pole_i_alt}
	\lim_{\dslash{p} \to m} \left(\dslash{p} - m \right) \frac{1}{Z^{2}(\rho)}\langle \Omega^{(\rho)} | T \hat{\psi}(p)\bar{\hat{\psi}}(-p) | \Omega^{(\rho)} \rangle = i.
\end{align}
Here $\hat{\psi}(p)$ denotes the Fourier transform of $\psi(x)$, and $Z(\rho)$ is a factor that depends on the exact structure of $1PI$ vertex function of $F(\psi(x),\bar{\psi}(x))$ and $\psi(x)$. The demonstration of this formula is based on the spectral properties of $1PI$ vertex functions. The generalization to the fermionic field is immediate and we refer to \cite{Bergere:1975tr} for details, as well as the explicit formulas in the appendix \ref{sec:appendix} for the fermionic case. We give below another, simpler derivation of the formula \eqref{smatrix:pole_i_alt}, taking into account the choice of a false vacuum \eqref{smatrx:pseudo_vacuum_original_theory} for the original theory. First we address the relation of the vacuum $\Omega^{(\rho)}$ of the transformed theory to the pseudo-vacuum of the original theory $\Omega^{(0)}$ \eqref{smatrx:pseudo_vacuum_original_theory}. For the sake of clarity, we change the notations here and denote the field of the original theory by $\chi(x)$, and the field of the transformed theory by $\psi(x)$. Thus, in this notation, the field transformations are: 
\begin{align}
	\chi(x)=\psi(x)+\rho F(\psi,\bar{\psi}), \label{smatrix:chi_transform_a} \\
	\bar{\chi}(x)=\bar{\psi}(x)+\rho \bar{F}(\psi,\bar{\psi}),\label{smatrix:chi_transform_b}
\end{align}
and the choice of the false vacuum \eqref{smatrx:pseudo_vacuum_original_theory} takes the form:
\begin{equation}
	\chi(x)|\Omega^{(0)}\rangle =0. \label{smatrx:pseudo_vacuum_original_theory_chi}
\end{equation}
The original Lagrangian takes the form $\mathscr{L}(\chi,\bar{\chi})$, and the transformed Lagrangian becomes $\mathscr{L}_{\rho}(\psi,\bar{\psi}) \equiv \mathscr{L}(\chi(\psi,\bar{\psi}),\bar{\chi}(\psi,\bar{\psi}))$. One would like, for the reasons explained above, to be able to choose a pseudo-vacuum for the transformed Lagrangian $\mathscr{L}_{\rho}(\psi,\bar{\psi})$, which now should satisfy the condition:
\begin{equation}
	\psi(x)|\Omega^{(\rho)}\rangle =0. \label{smatrx:pseudo_vacuum_transformed_theory}
\end{equation}
We must, however, show that this is possible and consistent with the field redefinitions \eqref{smatrix:chi_transform_a} and \eqref{smatrix:chi_transform_b}. Indeed, as we had noted in section \ref{sec:aafov}, there exists an inverse transformation to \eqref{smatrix:chi_transform_a},\footnote{For its explicit form see appendix {\bf D} of \cite{Alday:2005jm}} and, therefore, one can write the inverse functions $\psi(\chi,\bar{\chi})$, and $\bar{\psi}(\chi,\bar{\chi})$. In fact these functions have a polynomial form which extend up to the seventh order in $\chi(x)$, and $\bar{\chi}(x)$, as well as their derivatives. We then find:
\begin{equation}
	\psi(x) |\Omega^{(0)}\rangle = \Big( \chi(x) - \rho G(\chi,\bar{\chi})(x)\Big) |\Omega^{(0)}\rangle , \label{smatrix:chi_psi_vacuum}
\end{equation}
where $G(\chi,\bar{\chi})$ is some polynomial in $\chi(x)$, $\bar{\chi}(x)$ and their derivatives. Thus, up to the normal ordering for the operators and some regularization of the singular operator product that appears in in $G(\chi,\bar{\chi})$,\footnote{This point will be discussed in more details in the next section.} we find:
\begin{equation}
	\psi(x) |\Omega^{(0)}\rangle  = 0. \label{smatrix:psi_on_Omega_0}
\end{equation}
Thus, the vacuum $\Omega^{(\rho)}$, corresponding to the Lagrangian $\mathscr{L}_{\rho}(\psi,\bar{\psi})$, should satisfy the relation:
\begin{equation}
	| \Omega^{(\rho)} \rangle = Z(\rho)| \Omega^{(0)} \rangle,\label{smatrix:vacuum_renormalization}
\end{equation}
where $Z(\rho)$ is some renormalization constant depending in general on the parameter $\rho$, and which should satisfy the condition $Z(0)=1$. Hence, we have shown that $\Omega^{(\rho)}$ is a false vacuum \eqref{smatrx:pseudo_vacuum_transformed_theory} of the transformed theory with Lagrangian $\mathscr{L}_{\rho}(\psi,\bar{\psi})$.

This relation will be confirmed in the next section where we will show that the quantum Hamiltonian can be diagonalized in the vacuum defined by \eqref{smatrix:vacuum_renormalization} and reproduces the $S$-matrix of the $AAF$ model.

Next, we note that the free part of the Lagrangian $\mathscr{L}_{\rho}$ coincides with that of the Lagrangian of the original theory \eqref{aafov:lag_relativistic}. Since we have shown above that the vacuum of the transformed theory can be chosen to be the pseudo-vacuum $|\Omega^{(\rho)}\rangle$ \eqref{smatrx:pseudo_vacuum_transformed_theory}, we can use the non-renormalizability theorem of \cite{Klose:2006dd} to show that the full propagator  $\langle \Omega^{(\rho)}| T \psi(x) \bar{\psi}(y)| \Omega^{(\rho)}\rangle$ coincides with the free propagator \eqref{smatrix:propagator_free}. We find:
\begin{align}
	\label{smatrix:propagator_free_rho} \frac{1}{Z^{2}(\rho)}\langle \Omega^{(\rho)}| T \psi(x) \bar{\psi}(y)| \Omega^{(\rho)}\rangle = \langle \Omega^{(0)}| T \psi(x) \bar{\psi}(y)| \Omega^{(0)}\rangle = \int \frac{d^2p}{4 \pi^2}\: D(p) e^{-ip \cdot (x-y)},
\end{align}
where $D(p)$ is the free propagator in the momentum space:
\begin{equation} \label{smatrix:propagator_free_momentum}
	D(p) = \frac{i \left(\dslash{p} + m \right)}{p^2 - m^2 + 2 i \epsilon p_0}.
\end{equation}
Writing this formula in momentum space and evaluating the residue of both sides at the simple pole $\dslash{p}=m$, we find the formula \eqref{smatrix:pole_i_alt}. This simple derivation of the formula \eqref{smatrix:pole_i_alt} reproduces the result in \cite{Bergere:1975tr} which was obtained in the \emph{physical} vacuum and with the use of the spectral properties of the $1PI$ interaction vertices. Here we have avoided those complications and the additional assumption that there is a simple pole at $\dslash{p}=m$, by using the false vacuum and the non-renormalization theorem of \cite{Klose:2006dd}.

Using the formula \eqref{smatrix:pole_i_alt}, the $S$-matrix can be found from the reduction formula \cite{Itzykson:1980rh} written in terms of the renormalized fields $\psi^{r} \equiv \frac{1}{Z(\rho)}\psi$:\footnote{The antiparticle states do not enter in the reduction formula \eqref{smatrix:LSZ}, due to non-renromalization theorem discussed earlier.}
\begin{align}\label{smatrix:LSZ}
& S(k_{1},\ldots,k_{n};\: q_{1},\ldots,q_{l};\rho) \\ \sim
&  \mathop{\lim_{\dslash{q}_{j} \to m}}_{\dslash{k}_{i} \to m} \bar{u}(q_{1})\left(\dslash{q}_{1} - m \right)\cdot \ldots \cdot \langle \Omega^{(\rho)}| T \frac{1}{Z(\rho)}\hat{\psi}^{r}(q_{1}) \ldots  \frac{1}{Z(\rho)}\bar{\hat{\psi}}^{r}(k_{n})| \Omega^{(\rho)}\rangle \cdot \ldots \cdot\left(\dslash{k}_{n} - m \right)u(k_{n}) \nonumber
\end{align}
The independence of $S(q_{1},\ldots,q_{k};\: p_{1},\ldots,p_{n};\rho)$ on the parameter $\rho$ is a straightforward generalization, and can be proved without any changes following the same steps presented in \cite{Bergere:1975tr}, by using the corresponding formulas for the fermion case, collected in the appendix \ref{sec:appendix}.

\section{Diagonalization of $AAF$ Hamiltonian}\label{sec:diag}
 
The perturbative proof of the invariance of the $S$-matrix  suffers from several limitations. Firstly, as we stressed above, there is no proof so far that the $AAF$ model is a renormalizable model. Secondly, as emphasized in \cite{Sharatchandra:1978dj} the equivalence theorem may not be valid in the non-perturbative regime. Since the model under consideration is a classically integrable model, and is expected to be integrable for the quantum theory as well, we are mainly interested in the non-perturbative analysis. Integrability provides an opportunity to check and interrelate several fine points that appear in the proof of the equivalence theorem. For instance, it is interesting to see how the renormalization of the wave-functions in \eqref{smatrix:LSZ} emerges through more reliable methods of integrable systems.  There are essentially two approaches to  non-perturbative analysis: the inverse scattering method, and the direct diagonalization of the corresponding Hamiltonian via the coordinate Bethe ansatz. To realize the first program one needs to understand the quantization of non-ultralocal integrable models, which so far is an ill-understood and unsolved problem. There has been some progress in this direction \cite{Melikyan:2012kj,Melikyan:2014yma} for the non-ultralocal models that describe models such as the $AAF$ model, however the program is still far from being complete (see for example \cite{SemenovTianShansky:1995ha, Freidel:1991jv,Freidel:1991jx, Kundu:1996hb,Kundu:2003cu, Melikyan:2014temp} and the references therein). 

Thus, we choose here the second direction, which requires the direct diagonalization of the Hamiltonian of the $AAF$ model. As we will see below, in the process of diagonalization one encounters new types of difficulties, associated with the singular behavior of generalized functions. These singularities appear  as a result of the operators products at the same point, and cannot be avoided, as it is the case for some simpler integrable models. We refer the reader to the papers \cite{Melikyan:2008ab, Melikyan:2010fr} for a more detailed discussion and methods which were used in order to show the quantum integrability of the Landau-Lifshitz model directly in the continuous case. Since there is currently no lattice description of the $AAF$ model, one has to deal with such singularities directly in the continuous theory as well.

 
We briefly review the method used in \cite{Melikyan:2008ab, Melikyan:2010fr} for the Landau-Lifshitz model to deal with the singular operator products, and give a slightly different and more convenient formulation in terms of the \emph{Sklyanin's product} \cite{Sklyanin:1988}, in order to proceed with the diagonalization of the $AAF$ Hamiltonian. The Hamiltonian of the $AAF$ model can be easily read off from the transformed Lagrangian \eqref{aafov:lagrang_transformed} and has the form:
\begin{align}
\mathcal{H} & =-\frac{i}{2}\left(\psi_{i_{1}}^{\dagger}\gamma_{i_{1}i_{2}}^{3}\partial_{1}\psi_{i_{2}}-\partial_{1}\psi_{i_{1}}^{\dagger}\gamma_{i_{1}i_{2}}^{3}\psi_{i_{2}} \right)+m\psi_{i_{1}}^{\dagger}\gamma_{i_{1}i_{2}}^{0}\psi_{i_{2}}\nonumber \\
 & \qquad+\frac{i}{2}g_{2}\left(\psi_{i_{1}}^{\dagger}\psi_{j_{1}}^{\dagger}\gamma_{i_{1}i_{2}}^{0}\gamma_{j_{1}j_{2}}^{3}\psi_{i_{2}}\partial_{1}\psi_{j_{2}}-\psi_{i_{1}}^{\dagger}\partial_{1}\psi_{j_{1}}^{\dagger}\gamma_{i_{1}i_{2}}^{0}\gamma_{j_{1}j_{2}}^{3}\psi_{i_{2}}\psi_{j_{2}} \right)\nonumber \\
 & \qquad+\frac{g_{2}}{2m} \left(\psi_{i_{1}}^{\dagger}\psi_{j_{1}}^{\dagger}\gamma_{i_{1}i_{2}}^{0}\gamma_{j_{1}j_{2}}^{0}\partial_{1}\psi_{i_{2}}\partial_{1}\psi_{j_{2}}+\partial_{1}\psi_{i_{1}}^{\dagger}\partial_{1}\psi_{j_{1}}^{\dagger}\gamma_{i_{1}i_{2}}^{0}\gamma_{j_{1}j_{2}}^{0}\psi_{i_{2}}\psi_{j_{2}} \right)\nonumber \\
 & \qquad-\left(\frac{g_{3}+2g_{2}^{2}}{8m}\right) \left(\psi_{i_{1}}^{\dagger}\psi_{j_{1}}^{\dagger}\partial_{1}\psi_{k_{1}}^{\dagger}\gamma_{i_{1}i_{2}}^{0}\gamma_{j_{1}j_{2}}^{0}\gamma_{k_{1}k_{2}}^{0}\psi_{i_{2}}\psi_{j_{2}}\partial_{1}\psi_{k_{2}} \right)\nonumber \\
 & \qquad+i\frac{g_{2}^{2}}{8m^{2}}\left(\psi_{i_{1}}^{\dagger}\psi_{j_{1}}^{\dagger}\partial_{1}\psi_{k_{1}}^{\dagger}\gamma_{i_{1}i_{2}}^{0}\gamma_{j_{1}j_{2}}^{0}\gamma_{k_{1}k_{2}}^{3}\psi_{i_{2}}\psi_{j_{2}}\partial_{1}^{2}\psi_{k_{2}} -\psi_{i_{1}}^{\dagger}\psi_{j_{1}}^{\dagger}\partial_{1}^{2}\psi_{k_{1}}^{\dagger}\gamma_{i_{1}i_{2}}^{0}\gamma_{j_{1}j_{2}}^{0}\gamma_{k_{1}k_{2}}^{3}\psi_{i_{2}}\psi_{j_{2}}\partial_{1}\psi_{k_{2}} \right)\nonumber \\
 & \qquad-i\frac{g_{2}^{2}}{2m^{2}}\left(\psi_{i_{1}}^{\dagger}\psi_{j_{1}}^{\dagger}\partial_{1}\psi_{k_{1}}^{\dagger}\gamma_{i_{1}i_{2}}^{0}\gamma_{j_{1}j_{2}}^{0}\gamma_{k_{1}k_{2}}^{3}\psi_{i_{2}}\partial_{1}\psi_{j_{2}}\partial_{1}\psi_{k_{2}}-\psi_{i_{1}}^{\dagger}\partial_{1}\psi_{j_{1}}^{\dagger}\partial_{1}\psi_{k_{1}}^{\dagger}\gamma_{i_{1}i_{2}}^{0}\gamma_{j_{1}j_{2}}^{0}\gamma_{k_{1}k_{2}}^{3}\psi_{i_{2}}\psi_{j_{2}}\partial_{1}\psi_{k_{2}}\right)\nonumber \\
 & \qquad+\frac{g_{2}^{3}}{2m^{3}}\left(\psi_{i_{1}}^{\dagger}\psi_{j_{1}}^{\dagger}\partial_{1}\psi_{k_{1}}^{\dagger}\partial_{1}\psi_{l_{1}}^{\dagger}\gamma_{i_{1}i_{2}}^{0}\gamma_{j_{1}j_{2}}^{0}\gamma_{k_{1}k_{2}}^{0}\gamma_{l_{1}l_{2}}^{0}\psi_{i_{2}}\psi_{j_{2}}\partial_{1}\psi_{k_{2}}\partial_{1}\psi_{l_{2}}\right).\label{opreg:hamiltonian}
\end{align}
Is quite impressive that such a complex Hamiltonian describes an integrable model.\footnote{Its worth mentioning that the Hamiltonian of the original theory \eqref{aafov:lag_relativistic} has a very simple form. In contrast, the Dirac brackets structure is a very complicated one, while for the transformed theory with the Hamiltonian \eqref{opreg:hamiltonian} one has the standard Poisson brackets.} The Hamiltonian is very complicated in comparison to, e.g., the fermionic Thirring model, which has a much simpler Hamiltonian.

The simplicity of the fermionic Thirring model is not only due to the presence of the higher order terms in fermionic fields in \eqref{opreg:hamiltonian}. The main difficulty arises due to the presence of derivatives in the interaction terms. Its easy to check that any naive attempt to diagonalize the Hamiltonian \eqref{opreg:hamiltonian} immediately produces singular terms of the form $ \sim \partial_{1}^{2}\delta(0)$. For some simpler models with delta-function potential, e.g., the fermionic Thirring model, this problem is usually avoided and is dealt with in the standard manner. This is, however, not correct in general, and even for delta-function potentials it is necessary to construct the self-adjoint extensions. For the singularities that are exhibited in the $AAF$ model Hamiltonian, which are $ \sim \partial_{1}^{2}\delta(0)$, the problem is even more severe, and one cannot avoid performing a more careful analysis. 

First we introduce a regularization of the operator products in \eqref{opreg:hamiltonian}. This is done by means of Sklyanin's product, which for the product of two operators has the form \cite{Sklyanin:1988}:\footnote{Alternatively, one could proceed as in \cite{Melikyan:2008ab, Melikyan:2010fr} and introduce regularized fields:
\begin{align}
	\psi_{\mathcal{F}} (x) = \int dy F_{\eta}(x-y) \psi(y),\label{opreg:old_F_reg}
\end{align}
where $F_{\eta}(x-y)$ is some smooth symmetric function which depends on $\eta$ in such a way that in the limit $\eta \to 0$ one has : $F_{\eta}(x-y) \leadsto \delta(x-y)$. It can be shown that both methods lead to the same results. We will use in this paper the regularization based on a more transparent Sklyanin's product.}
\begin{equation}
A(x) \circ B(x)
\equiv \lim_{\Delta \to 0} \frac{1}{\Delta^{2}}\int_{x-\Delta/2}^{x+\Delta/2}du\int_{x-\Delta/2}^{x+\Delta/2}dv A(u)B(v). \label{opreg:sklyanin_prod}
\end{equation}
For a product of $n$ operators $A_{1}(x),\ldots, A_{n}(x)$ the generalization of the above product is straightforward:
\begin{equation}
A_{1}(x) \circ \ldots \cdot \circ A_{n}(x)
\equiv \lim_{\Delta \to 0} \frac{1}{\Delta^{n}}\int_{x-\Delta/2}^{x+\Delta/2}du_{1} \ldots \int_{x-\Delta/2}^{x+\Delta/2}du_{n} A_{1}(u_{1}) \cdot \ldots \cdot A_{n}(u_{n}). \label{opreg:sklyanin_prod_general}
\end{equation}
Thus, our starting point is the Hamiltonian in \eqref{opreg:hamiltonian} where instead of the usual product, resulting in ill-defined singular expressions, we use the Sklyanin's product \eqref{opreg:sklyanin_prod_general}. We denote the Hamiltonian  \eqref{opreg:hamiltonian}  regularized in such a way by $\mathcal{H}_{\Delta}$. All the following computations will be performed with this Hamiltonian and the limit $\Delta \to 0$ will be taken only at the end. We take the vacuum $|0\rangle$ to be the same pseudo-vacuum considered in the perturbative analysis in \cite{Klose:2006dd, Melikyan:2011uf} and in section \ref{sec:eq}, i.e., the vacuum satisfies the condition \eqref{smatrx:pseudo_vacuum_transformed_theory}:\footnote{To simplify the notation, in this section we denote: $|0\rangle \equiv |\Omega^{(\rho)} \rangle$, for $\rho=1$.} $\psi |0 \rangle = 0$. Then it is clear from \eqref{opreg:hamiltonian} that $\mathcal{H}_{\Delta} |0 \rangle = 0$. 

For the one-particle sector that Hamiltonian coincides with the free theory, and, thus, we consider the state:
\begin{equation}
\left|\psi\right\rangle _{1}=\int dx\psi_{k}^{\dagger}(x)\chi^{k}(x)\left|0\right\rangle, \label{opreg:1_particle_state}
\end{equation}
which is the eigenstate of the Hamiltonian $\mathcal{H}_{\Delta}$, i.e., $\mathcal{H}_{\Delta} \left|\psi\right\rangle _{1} = E_{1}\left|\psi\right\rangle _{1}$, provided the quantum-mechanical wave function $\chi^{k}(x)$ has the form:
\begin{equation}
\chi^{k}(x)=\left(\begin{array}{c}
e^{\theta/2}\\
e^{-\theta/2}
\end{array}\right)e^{ix\, m\, \sinh(\theta)},
\end{equation}
and eigenvalue $E_{1}=m\, \cosh(\theta)$, where $\theta$ is the rapidity.

Let us now consider the first non-trivial case - the two-particle sector. 
One can try to proceed as in the case of the fermionic Thirring model and consider the naive generalization of the form:
\begin{equation}
\left\vert \psi\right\rangle _{2}=\int dxdy\psi_{k_{1}}^{\dagger}(x)\psi_{k_{2}}^{\dagger}(y)\chi^{k_{1}k_{2}}(x,y)\left\vert 0\right\rangle \label{2p_wave_fun}
\end{equation}
Here, the two-particle quantum-mechanical wave-function $\chi^{k_{1}k_{2}}(x,y)$ should satisfy the anti-symmetry condition: 
\begin{equation}
	\chi^{k_{1}k_{2}}(x,y)=-\chi^{k_{2}k_{1}}(y,x).\label{opreg:two_part_wf_wrong}	
\end{equation}
However, even in the case of the fermionic Thirring model, the corresponding functions $\chi^{k_{1}k_{2}}(x,y)$ are not continuous on the line $x=y$. In the $AAF$ model case we expect that neither the function $\chi^{k_{1}k_{2}}(x,y)$ nor its derivatives will be continuous functions on the line $x=y$. Thus, instead of the ill-defined  expression \eqref{opreg:two_part_wf_wrong}on the line $x=y$, we take the two-particle state to be the well-defined principal value integral:
\begin{equation}
\left\vert \psi\right\rangle _{2}=\lim_{\eta\rightarrow0}\overset{+\infty
}{\underset{-\infty}{\int}}dy\left(  \overset{y-\eta}{\underset{-\infty}{\int
dx}}+\overset{+\infty}{\underset{y+\eta}{\int}}dx\right)  \psi_{k_{1}}
^{\dagger}(x)\psi_{k_{2}}^{\dagger}(y)\chi^{k_{1}k_{2}}(x,y)\left\vert 0\right\rangle
\label{opreg:two_part_wf_pv}
\end{equation}
This choice immediately raises the question of what happens on the line $x=y$. Even though the function $\chi^{k_{1}k_{2}}(x,y)$ is not continuous, one could still write an additional two-particle wave function for the points on the line $x=y$. Later we will see that such an extra term will be required in order to diagonalize the Hamiltonian $\mathcal{H}_{\Delta}$. We also note that a similar extra term appears in the Landau-Lifshitz model (see \cite{Sklyanin:1988,Melikyan:2008ab} and the formula \eqref{opreg:LL_2p} below).

After some lengthy calculations, where one has to take into account the fact that neither the wave-function $\chi^{k_{1}k_{2}}(x,y)$ nor its derivatives are in general continuous functions, the action of the regularized Hamiltonian $\mathcal{H}_{\Delta}$  on $\left\vert
\psi\right\rangle _{2}$ in \eqref{opreg:two_part_wf_pv} can be organized in powers of the regularization parameter $\Delta$ as follows:\footnote{The parameter $\eta$ is the regulator of the
principal value prescription in \eqref{opreg:two_part_wf_pv}, and the parameter $\Delta$ is the regulator of $\mathcal{H}_{\Delta}$. In the course of the calculation we will always consider first the limit $\eta\rightarrow0$, and only after that we remove the regularization $\Delta \rightarrow0$.}
\begin{align}
H_{\Delta}\left\vert \psi\right\rangle _{2} &=m[\cosh(\theta_{1})+\cosh(\theta
_{2})]\left\vert \psi\right\rangle _{2}\nonumber\\ &+\Delta^{0}\int dy\Omega
_{1}(y)\psi_{{1}}^{\dagger}(y)\psi_{{2}}^{\dagger}(y)\left\vert 0\right\rangle+\frac{1}{\Delta}\int dy\Omega_{2}(y)\psi_{{1}}^{\dagger}(y)\psi_{{2}}^{\dagger}(y)\left\vert 0\right\rangle +\Delta\int dy\Omega_{3}(y).\label{opreg:H_on_psi2_main}
\end{align}
Thus, the Hamiltonian can be diagonalized provided the last three integrals, containing $\Omega_{1,}\Omega_{2}$ and $\Omega_{3}$ vanish.
The exact expression for $\Omega_{1}$ has the following form:
\begin{align}
\Omega_{1}  & =2i\left[  \chi^{12}(y-\eta,y)-\chi^{12}(y+\eta,y)\right]
+2i\left[  \chi^{21}(y-\eta,y)-\chi^{21}(y+\eta,y)\right]  \label{opreg:Omega1} \\
&- ig_{2}\left[  \left.  \partial_{x}\chi^{11}(x,y)\right\vert _{x=y-\eta
}+\left.  \partial_{x}\chi^{11}(x,y)\right\vert _{x=y+\eta}\right]
-ig_{2}\left[  \left.  \partial_{x}\chi^{22}(x,y)\right\vert _{x=y-\eta
}+\left.  \partial_{x}\chi^{22}(x,y)\right\vert _{x=y+\eta}\right] \nonumber\\
&+ \frac{1}{m}g_{2}\left[  \left.  \partial_{y}\partial_{x}\chi^{12}%
(x,y)\right\vert _{x=y-\eta}+\left.  \partial_{y}\partial_{x}\chi
^{12}(x,y)\right\vert _{x=y+\eta}\right] -\frac{1}{m}g_{2}\left[  \left.  \partial_{y}\partial_{x}\chi^{21}%
(x,y)\right\vert _{x=y-\eta}+\left.  \partial_{y}\partial_{x}\chi
^{21}(x,y)\right\vert _{x=y+\eta}\right],  \nonumber
\end{align}
while $\Omega_{2}$ has the form:
\begin{align}
\Omega_{2}  & =ig_{2}\left[\chi^{11}(y-\eta,y)-\chi^{11}(y+\eta,y)\right]+ig_{2}
\left[\chi^{22}(y-\eta,y)-\chi^{22}(y+\eta,y)\right]\label{opreg:Omega2}\\
& +\frac{1}{m}g_{2}\left[\left.  \partial_{x}\chi^{12}(x,y)\right\vert _{x=y-\eta
}-\left.  \partial_{x}\chi^{12}(x,y)\right\vert _{x=y+\eta}\right]-\frac{1}{m}
g_{2}\left[\left.  \partial_{x}\chi^{21}(x,y)\right\vert _{x=y-\eta}-\left.
\partial_{x}\chi^{21}(x,y)\right\vert _{x=y+\eta}\right]\nonumber\\
& -\frac{1}{2m}g_{2}\left[\frac{d}{dy}\left(\chi^{12}(y-\eta,y)\right)-\frac{d}{dy}\left(\chi^{12}(y+\eta
,y)\right)\right]+\frac{1}{2m}g_{2}\left[\frac{d}{dy}\left(\chi^{21}(y-\eta,y)\right)-\frac{d}{dy}\left(\chi^{21}(y+\eta
,y)\right)\right]\nonumber
\end{align}
The exact expression for $\Omega_{3}$, corresponding to the solution of the equations $\Omega_{1}=0$ and $\Omega_{2}=0$ will be given below (see \eqref{opreg:Omega3_ansatz}).

Our first result is that the equation $\Omega_{1}=0$ reproduces exactly the two-particle $S$-matrix of the $AAF$ model \eqref{aafov:smatrix}. To show this, we start with the standard coordinate Bethe ansatz for the wave-function of an integrable model \cite{Korepin:1997bk}:
\begin{align}
\text{For }x  & <y:\text{ }\chi^{k_{1}k_{2}}(x,y)=Au^{k_{1}}(\theta
_{1})u^{k_{2}}(\theta_{2})e^{i(p_{1}x+p_{2}y)}-Bu^{k_{1}}(\theta_{2})u^{k_{2}
}(\theta_{1})e^{i(p_{1}y+p_{2}x)}\label{opreg:ansatz_a}\\
\text{For }x  & >y:\text{ }\chi^{k_{1}k_{2}}(x,y)=Bu^{k_{1}}(\theta
_{1})u^{k_{2}}(\theta_{2})e^{i(p_{1}x+p_{2}y)}-Au^{k_{1}}(\theta_{2})u^{k_{2}
}(\theta_{1})e^{i(p_{1}x+p_{2}y)}\label{opreg:ansatz_b}
\end{align}
Here we have denoted $u^{k_{1}}(\theta_{1})=u_{0}\left(
\begin{array}
[c]{c}
e^{\theta/2}\\
e^{-\theta/2}
\end{array}
\right)$, where $u_{0}$ is a normalization constant, and the momentum is $p_{i}
=m\sinh(\theta_{i})$.
The above ansatz is suitable for a Lorentz invariant theory, e.g., for the fermionic Thirring model. Since the transformation of the fields \eqref{aafov:trans_psi} results in a Lagrangian that is not invariant under the Lorentz transformation, the form of \eqref{opreg:ansatz_a}, \eqref{opreg:ansatz_b} can now be justified by invoking the equivalence theorem (see the discussion after \eqref{smatrix:simple_pole}). Thus, we expect that the form of the wave-function \eqref{opreg:ansatz_a}, \eqref{opreg:ansatz_b} is the same as in the original Lorentz invariant theory \eqref{aafov:lag_relativistic}. Indeed, substituting the formulas \eqref{opreg:ansatz_a} and \eqref{opreg:ansatz_b} into the equation $\Omega_{1}=0$ \eqref{opreg:Omega1}, one finds after some straightforward calculations that the  $S$-matrix has the form:
\begin{equation}
S(\theta_{1},\theta_{2})=\frac{1-i\frac{mg_{2}}{4}\sinh(\theta_{1}-\theta_{2})}{1+i\frac{mg_{2}}%
{4}\sinh(\theta_{1}-\theta_{2})}.\label{opreg:smatrix}
\end{equation}
This is exactly the two-particle $S$-matrix of the $AAF$ model \eqref{aafov:smatrix}, obtained from perturbative calculations in the original theory. 

We must still deal with the extra terms $\Omega_{2}$ and $\Omega_{3}$ in \eqref{opreg:H_on_psi2_main}. Substituting the ansatz \eqref{opreg:ansatz_a}, \eqref{opreg:ansatz_b} into \eqref{opreg:Omega2}, one obtains:
\begin{align}\label{opreg:Omega2_ansatz}
&\int dy  \Omega_{2}(y)\psi_{{1}}^{\dagger}(y)\psi_{{2}}^{\dagger}(y)\left\vert 0\right \rangle \nonumber \\ 
&=2(A-B)ig_{2}\left[  \cosh(\theta_{1})+\cosh(\theta_{2})\right] \cosh(\frac{\theta_{1}-\theta_{2}}{2})\int
dy\psi_{1}^{\dagger}(y)\psi_{2}^{\dagger}(y)e^{i(p_{1}+p_{2})y}\left\vert 0\right\rangle,
\end{align}
Thus, $\Omega_{2}$ does not vanish, except for an uninteresting non-scattering case, and the corresponding term in \eqref{opreg:H_on_psi2_main} goes to infinity as $\Delta \rightarrow 0$. We also note here that using the formulas \eqref{opreg:ansatz_a} and \eqref{opreg:ansatz_b} one can show that the term $\sim \Omega_{3}$ in \eqref{opreg:H_on_psi2_main} takes the following form:
\begin{align}
\int dy\Omega_{3}(y)  &=
-ic_{1}g_{2}\int dy\left[  \psi_{1}
^{\dagger}(y)\partial_{y}\psi_{1}^{\dagger}(y)e^{i(p_{1}+p_{2})y}\right]
 -ic_{1}g_{2}\int dy\left[  \psi_{2}^{\dagger}(y)\partial_{y}\psi_{2}^{\dagger}
(y)e^{i(p_{1}+p_{2})y}\right] \nonumber \\
& +c_{1}\frac{2}{m}g_{2}\int dy\left[  \partial_{y}\psi_{1}^{\dagger}(y)\partial_{y}
\psi_{2}^{\dagger}(y)e^{i(p_{1}+p_{2})y}\right],  \label{opreg:Omega3_ansatz}
\end{align}
where the constant $c_{1}$ is given by the formula $ c_{1}= 2(A+B)\sinh(\frac{\theta_{1}-\theta_{2}}{2})$. Hence, the ansatz \eqref{opreg:two_part_wf_pv} is not enough in this case to diagonalize the Hamiltonian. To solve this problem, we recall that in our ansatz \eqref{opreg:two_part_wf_pv} for the wave-function $\chi^{k_{1}k_{2}}(x,y)$ we had removed the points corresponding to the line $x=y$, since we expect that neither the wave function  nor its derivatives are continuous functions. One, therefore, should add a state corresponding to the removed line $x=y$. Clearly, the only term that one can form from the components $\psi_{1}(x), \psi_{2}(x)$ of a fermion field is a state of the type:
\begin{equation}\label{opreg:missed_term_naive}
	\int dy\psi_{1}
	^{\dagger}(y)\psi_{2}^{\dagger}(y)f(y) 
\end{equation}
where $f(y)$ is some continuous function. The formula \eqref{opreg:missed_term_naive} should be understood in the sense of the regularization (see \eqref{opreg:sklyanin_prod} and \eqref{opreg:sklyanin_prod_general}) which we have adopted for a product of operators at the same point. Thus, the product of $\psi_{1}^{\dagger}(y)\psi^{\dagger}_{2}(y)$ is regularized using the Sklyanin's product, and, written explicitly, the formula \eqref{opreg:missed_term_naive} can be cast in the following form:
\begin{equation}
\left\vert \psi^{\prime}\right\rangle _{2}=\frac{1}{\eta^{2}}\overset{+\infty
}{\underset{-\infty}{\int}}dv\overset{v+\eta/2}{\underset{v-\eta/2}{\int}
}dx\overset{v+\eta/2}{\underset{v-\eta/2}{\int}}dy\left[  \psi_{k_{1}}
^{\dagger}(x)\psi_{k_{2}}^{\dagger}(y)\right]  f^{k_{1}k_{2}}(x,y)\left\vert
0\right\rangle. \label{opreg:psi2_extra}
\end{equation}
In this expression we consider the functions $f^{k_{1}k_{2}}(x,y)$ to be continuous. It is then clear that in the limit $\eta \rightarrow 0$ the terms $\sim \psi^{\dagger}_{1}(y) \psi^{\dagger}_{1}(y)$ and $\sim \psi^{\dagger}_{2}(y) \psi^{\dagger}_{2}(y)$ will disappear and one obtains the ill-defined expression \eqref{opreg:missed_term_naive}. In contrast, there are no singular operator products in \eqref{opreg:psi2_extra}, and the expression is well-defined. Thus, we take our complete two-particle wave function to be:
\begin{equation}\label{opreg:complete_wave_function}
	\left\vert
	\psi_{complete}\right\rangle _{2}=\left\vert \psi\right\rangle _{2}+\left\vert
	\psi^{\prime}\right\rangle _{2}
\end{equation}
In section \ref{sec:sa} we will show that the addition of the extra term \eqref{opreg:psi2_extra}, and the constraints we will obtain below in order to diagonalize the Hamiltonian, correspond to the construction of self-adjoint extensions for the quantum-mechanical Hamiltonian, in a manner similar to the case of the Landau-Lifshitz model considered in details in \cite{Melikyan:2008ab,Melikyan:2010fr}. 

To show this, it is convenient to split the result of $H_{F}\left\vert \psi^{\prime}\right\rangle _{2}$ into two terms $H_{F}\left\vert \psi^{\prime}\right\rangle _{2} = \Lambda_{1} + \Lambda_{2}$.
The explicit form of the first term is:
\begin{align}\label{opreg:Lambda1}
\Lambda_{1}  & =  (-i){g_{2}}\int dy\left[  \psi_{1}^{\dagger}(y)\partial_{y}\psi_{1}^{\dagger}
(y)f^{12}(y,y)\right] \\
& -i{g_{2}}\int dy\left[  \psi_{2}^{\dagger}(y)\partial_{y}\psi_{2}^{\dagger}
(y)f^{12}(y,y)\right]  +2\frac{g_{2}}{m}\int dy\left[  \partial_{y}\psi_{1}^{\dagger}(y)\partial_{y}
\psi_{2}^{\dagger}(y)f^{12}(y,y)\right]  \nonumber
\end{align}
Comparing the above expression with \eqref{opreg:Omega3_ansatz} we see that when considering the action of the Hamiltonian $\mathcal{H}$ on the complete state \eqref{opreg:complete_wave_function} the term $\sim \Omega_{3}$ will cancel with the term $\Lambda_{1}$ above, if one sets:
\begin{equation}
f^{12}(y,y)=(-c_{1})e^{i(p_{1}+p_{2})y}\label{opreg:f12_y_y}
\end{equation}
In the next section we will also show that this choice of $f^{12}(y,y)$ is in perfect agreement with construction of self-adjoint extensions. Collecting the rest of the terms, we find:
\begin{align}\label{opreg:main1}
H_{F}\left\vert \psi_{complete}\right\rangle _{2}  & =m[\cosh(\theta_{1}
)+\cosh(\theta_{2})]\left\vert \psi_{2}\right\rangle_{2} \\
& +\frac{2}{\Delta}(A-B)ig_{2}\left[  \cosh(\theta_{1}
)+\cosh(\theta_{2})\right]  \cosh(\frac{\theta_{1}-\theta_{2}}{2})\int
dy\psi_{1}^{\dagger}(y)\psi_{2}^{\dagger}(y)e^{i(p_{1}+p_{2})y}\left\vert 0\right\rangle
\nonumber\\
& -ig_{2}\int dy\psi_{1}^{\dagger}(y)\psi_{2}^{\dagger}(y)\left[  \partial
_{x}f^{11}(x,y)+\partial_{x}f^{22}(x,y)\right]  |_{x=y}\left\vert
0\right\rangle \nonumber\\
& +\frac{2}{m}g_{2}\int dy\psi_{1}^{\dagger}(y)\psi_{2}^{\dagger}(y)\left[
\partial_{x}\partial_{y}f^{12}(x,y)\right]  |_{x=y}\left\vert 0\right\rangle
\nonumber\\
& -2i\int dy\psi_{1}^{\dagger}(y)\psi_{2}^{\dagger}(y)\left[  \partial_{x}
f^{12}(x,y)-\partial_{y}f^{12}(x,y)\right]  |_{x=y}\left\vert 0\right\rangle
\nonumber
\end{align}
It is easy to see from the above formula that the Hamiltonian can be diagonalized by choosing the functions $f^{11}(x,y)$, $f^{22}(x,y)$, and $f^{12}(x,y)$  as follows:
\begin{align}
f^{12}(x,y)  & =-(A+B)\Delta\left[  \lambda_{1}e^{i(p_{1}x+p_{2}
y)}-\lambda_{2}e^{i(p_{1}y+p_{2}x)}\right], \label{opreg:choice_f_12}\\
f^{11}(x,y)  & =-(A+B)\Delta a_{11}\left[  e^{i(p_{1}x+p_{2}
y)}-e^{i(p_{1}y+p_{2}x)}\right], \label{opreg:choice_f_11}\\
f^{22}(x,y)  & =-(A+B)\Delta a_{22}\left[  e^{i(p_{1}x+p_{2}
y)}-e^{i(p_{1}y+p_{2}x)}\right],\label{opreg:choice_f_22}
\end{align}
where $a_{11}, a_{22}$ are some constants, and $\lambda_{1}$ and $\lambda_{2}$ satisfy the condition:
\begin{equation}
	\lambda_{1}-\lambda
	_{2}=2\sinh(\frac{\theta_{1}-\theta_{2}}{2}).\label{opreg:lambda1_lambda2}
\end{equation}
This is consistent with the general antisymmetry property: $f^{ij}
(x,y)=-f^{ji}(y,x)$ and our choice for  $f^{12}(y,y)$ in \eqref{opreg:f12_y_y}. Restricting the solution to the case $a_{11}=a_{22}=0$, one easily finds the solution for $\lambda_{1}$ and $\lambda_{2}$:
\begin{align}
\lambda_{1}  & =\frac{1}{2}\left[  1+2\sinh(\frac{\theta_{1}-\theta_{2}}
{2})+g_{2}h_{1}+(g_{2})^{2}h_{2}\right] \label{opreg:solution_lambda1}  \\
\lambda_{2}  & =\frac{1}{2}\left[  1-2\sinh(\frac{\theta_{1}-\theta_{2}}
{2})+g_{2}h_{1}+(g_{2})^{2}h_{2}\right],\label{opreg:solution_lambda2}
\end{align}
where $h_{1}$ and $h_{2}$ are some functions of $\theta_{i}$ and
$\Delta$. Since the parameter $\Delta$ corresponds to the regularization of the Hamiltonian $\mathcal{H}_{\Delta}$, we conclude from the above formulas that the complete wave-function requires renormalization. This corresponds exactly to the perturbative proof of the equivalence theorem, and is the key point in proving the $S$-matrix invariance (see \eqref{smatrix:LSZ}). Thus, we can see now how the renormalization of the wave-function appears in the direct diagonalization and in the perturbative proof.

Finally, we show that:
\begin{equation}
	\mathcal{H}_{\Delta}\left\vert \psi_{complete}\right\rangle _{2} =m[\cosh(\theta_{1})+\cosh(\theta_{2})]\left\vert \psi_{complete}\right\rangle_{2},\label{opreg:Ham_diag_final}
\end{equation}
where the complete two-particle state has the form:
\begin{align}
	\left\vert \psi_{complete}\right\rangle _{2} &=\lim_{\eta\rightarrow0}\frac{1}{\eta^{2}}\overset{+\infty
}{\underset{-\infty}{\int}}dv\overset{v+\eta/2}{\underset{v-\eta/2}{\int}
}dx\overset{v+\eta/2}{\underset{v-\eta/2}{\int}}dy\left[  \psi_{k_{1}}
^{\dagger}(x)\psi_{k_{2}}^{\dagger}(y)\right]  f^{k_{1}k_{2}}(x,y)\left\vert
0\right\rangle\nonumber \\
&+ \lim_{\eta\rightarrow0}\overset{+\infty
	}{\underset{-\infty}{\int}}dy\left(  \overset{y-\eta}{\underset{-\infty}{\int
	dx}}+\overset{+\infty}{\underset{y+\eta}{\int}}dx\right)  \psi_{k_{1}}
	^{\dagger}(x)\psi_{k_{2}}^{\dagger}(y)\chi^{k_{1}k_{2}}(x,y)\left\vert 0\right\rangle. \label{opreg:2p_state_complete_final}
\end{align}
The two-particle $S$-matrix has the form \eqref{opreg:smatrix} and coincides with the expression obtained from the perturbative analysis of the original relativistic invariant theory \eqref{aafov:lag_relativistic}. We stress that these results are valid only in the pseudo-vacuum, in agreement with the proof of the $S$-matrix invariance given in the previous section. 

It is interesting to compare the form of the complete two-particle state \eqref{opreg:2p_state_complete_final} for the $AAF$ model with that of the Landau-Lifshitz model \cite{Sklyanin:1988,Melikyan:2008ab,Melikyan:2010fr}. Both models exhibit the same type of singularity in the quantum-mechanical Hamiltonian (namely, singularities of the type $\sim \delta''(0)$), and require construction of the self-adjoint extensions. The Hamiltonian for the isotropic  Landau-Lifshitz model has the form:
\begin{equation}
H=\frac{1}{2}\int dx \left( \partial_{x}\vec{S}\partial _{x}\vec{S}\right) \label{opreg:LL_hamiltonian}
\end{equation}
It was shown in \cite{Sklyanin:1988,Melikyan:2008ab} that the two-particle state can be written in the form:
\begin{equation} 
|\psi\rangle_{LL}  = \int 
dx \; g_{1}(x)\Psi_{2}^{\dagger}(x)   +\int \int_{x>y} dxdy \; g_{2}(x,y)
\Psi_{1}^{\dagger}(x)\Psi_{1}^{\dagger}(y) |0\rangle.  \label{opreg:LL_2p}
\end{equation}
The bosonic fields $\Psi_{n}(x)$ are defined by means of the following decompositions:
\begin{align} \label{opreg:cluster_decomposition}
S^{3}(x) & =s_{0}^{3}+\overset{\infty}{\underset{n=1}{\sum}} s_{n}^{3}\Psi_{n}^{\dagger}(x)\Psi_{n}(x)  \\
S^{+}(x) & =s_{0}^{\dagger}\Psi_{1}^{\dagger}(x)+\overset{\infty}{\underset{n=1}{\sum}} s_{n}^{3}\Psi_{n+1}^{\dagger}(x)\Psi_{n}(x)  \nonumber \\
S^{-}(x) & =s_{0}\Psi_{1}(x)+\overset{\infty}{\underset{n=1}{\sum}} s_{n}^{ \dagger 3}\Psi_{n}^{\dagger}(x)\Psi_{n+1}(x)  \nonumber
\end{align}
where $s_{0}^{3}=1; s_{0}^{\dagger}=\sqrt{2}; s_{n}^{3}=n; s_{n}^{\dagger}=\sqrt{(n+1)n}; (n\geq1)$, and satisfy the algebra:
\begin{align}
	\left[ \Psi_{m}(x),\Psi_{n}^{\dagger}(y)\right] &=\delta_{mn}\delta(x-y); \label{opreg:cluster_algebra} \\
	\Psi_{n}(x)|0\rangle &=0. \label{opreg:vacuum_cluster}
\end{align}
The functions $g_{1}(x)$ and $g_{2}(x,y)$ are continuous, and are defined as follows:
\begin{align}
\text{For }x  & <y:\text{ }g_{2}(x,y)=ce^{i(p_{1}x+p_{2}y)}+\bar{c}e^{i(p_{1}y+p_{2}x)}\label{opreg:ansatz_a_LL}\\
\text{For }x  & >y:\text{ }g_{2}(x,y)=\bar{c}e^{i(p_{1}x+p_{2}y)}+ce^{i(p_{1}y+p_{2}x)},\label{opreg:ansatz_b_LL}
\end{align}
and
\begin{equation}
	g_{1}(x)=g_{2}(x,x).\label{opreg:g1_g2_relation}
\end{equation}
The derivative of $g_{2}(x,y)$ is not a continuous function and satisfies the discontinuity condition:
\begin{equation}
	\left( \partial_{x} - \partial_{y} \right)g_{2}(x,y) \vert^{y=x+\eta}_{y=x-\eta} = \partial_{x} \partial_{y}g_{2}(x,y)\vert_{x=y}.\label{opreg:discontinuity_LL}
\end{equation}
Using the equations \eqref{opreg:ansatz_a_LL}-\eqref{opreg:discontinuity_LL}, one finds the $S$-matrix of the Landau-Lifshitz model:
\begin{equation}
S(p_{1,}p_{2})=\frac{2(p_{1}-p_{2})+ip_{1}p_{2}}{2(p_{1}-p_{2})}
\label{opreg:LL_S_matrix}
\end{equation}
It is clear now from the formulas \eqref{opreg:ansatz_a}, \eqref{opreg:ansatz_b} and \eqref{opreg:choice_f_12} that the form of the two-particle state \eqref{opreg:2p_state_complete_final} for the $AAF$ model  has the same structure as the corresponding state \eqref{opreg:LL_2p} for the Landau-Lifshitz model. Thus, the $AAF$ model can be considered as the fermionic counterpart of the bosonic Landau-Lifshitz model. The crucial difference, however, is that in the case of the Landau-Lifshitz model the function $g_{2}(x,y)$ is continuous at the line $x=y$, while in the case of the $AAF$ model the corresponding function $\chi^{k_{1}k_{2}}(x,y)$, together with its first derivative, are not continuous and satisfy a much more involved discontinuity condition $\Omega_{1}=0$, where $\Omega_{1}$ is given by the expression \eqref{opreg:Omega1}. As a consequence, the integrals in \eqref{opreg:2p_state_complete_final} are defined via the principal value. 

For the $n$-particle sector the analysis can be carried out in a similar manner, and will presented in a more complete form elsewhere. 
\section{Self-adjoint extensions}\label{sec:sa} 

We now interpret the results of the previous section by analyzing the self-adjointness of the quantum-mechanical Hamiltonian $\hat{H}$, corresponding to regularized Hamiltonian $\mathcal{H}_{\Delta}$. We show below that this condition requires the construction of self-adjoint extensions,\footnote{We refer the reader to the papers \cite{Melikyan:2008ab,Melikyan:2010fr} for a more detailed discussion of a similar analysis for the Landau-Lifshitz model.} and the resulting formulas reproduce the extra term in the two-particle state \eqref{opreg:psi2_extra}, required to diagonalize the Hamiltonian $\mathcal{H}_{\Delta}$.

We start by considering the vector space spanned by the elements of the form:\footnote{Here we again restrict our considerations for the case $n=2$. The general case is a straightforward generalization and will be presented elsewhere.}
\begin{equation}
\Psi=f_{1}^{ij}(x,y)\otimes f_{2}^{mn}(x,y),\label{sa:psi}
\end{equation}
where for notational convenience we have changed our notation: $\left( f^{ij}(x,y),\, \chi^{ij}(x,y) \right) \rightarrow \left(f_{1}^{ij}(x,y),\, f_{2}^{ij}(x,y) \right)$. Thus, the complete two-particle state \eqref{opreg:2p_state_complete_final} takes the form:
\begin{align}
	\left\vert \psi_{complete}\right\rangle _{2}  & =\overset{+\infty
	}{\underset{-\infty}{\int}}dy\left(  \overset{y-\eta}{\underset{-\infty}{\int
	dx}}+\overset{+\infty}{\underset{y+\eta}{\int}}dx\right)  \psi_{k_{1}}
	^{\dagger}(x)\psi_{k_{2}}^{\dagger}(y)f_{2}^{k_{1}k_{2}}(x,y)\left\vert 0\right\rangle \label{sa:psi_complete_a} \\
	&+\frac{1}{\eta^{2}}\overset{+\infty}{\underset{-\infty}{\int}}
	dv\overset{v+\eta/2}{\underset{v-\eta/2}{\int}}dx\overset{v+\eta
	/2}{\underset{v-\eta/2}{\int}}dy\left[  \psi_{k_{1}}^{\dagger}(x)\psi_{k_{2}}
	^{\dagger}(y)\right]  f_{1}^{k_{1}k_{2}}(x,y)\left\vert 0\right\rangle. \nonumber
\end{align}
We define the scalar product between two states of the type \eqref{sa:psi} as follows:
\begin{align}\label{sa:scalar_product}
	\langle \Phi \vert \Psi \rangle & =\alpha\left(
\frac{1}{\eta^{2}}\right)  \overset{+\infty}{\underset{-\infty}{\int}
}dv\overset{v+\eta/2}{\underset{v-\eta/2}{\int}}dx\overset{v+\eta
/2}{\underset{v-\eta/2}{\int}}dy\left[  g_{1}^{\ast ij}(x,y)f_{1}
^{ij}(x,y)\right]  \\
& +\overset{+\infty}{\underset{-\infty}{\int}}dy\left(  \overset{y-\eta
}{\underset{-\infty}{\int dx}}+\overset{+\infty}{\underset{y+\eta}{\int}
}dx\right)  g_{2}^{\ast ij}(x,y)f_{2}^{ij}(x,y),\nonumber
\end{align}
where $\Phi$ is a state defined similarly to \eqref{sa:psi}:
\begin{equation}
\Phi=g_{1}^{ij}(x,y)\otimes g_{2}^{mn}(x,y),\label{sa:phi}
\end{equation}
and $\alpha$ is come constant that we fix later by imposing the condition of the self-adjointness of $\hat{H}$.
The Hamiltonian action on $\Psi$ can be written in the following general form:
\begin{equation}
\hat{H}\Psi=\left(  \hat{h}f_{1}\right)  ^{ij}(x,y)\otimes\left(  \hat
{D}_{(x,y)}f_{2}\right)  ^{mn}(x,y)\label{sa:ham_act}%
\end{equation}
Here the operator $\left(  \hat{D}_{(x,y)}f_{2}\right)  ^{mn}$ has the form: 
\begin{equation}
\left(  \hat{D}_{(x,y)}f_{2}\right)  ^{mn}=\left[  \left(  -i\gamma_{mk_{1}}
^{3}\delta_{nk_{2}}\partial_{x}+m\gamma_{mk_{1}}^{0}\delta_{nk_{2}}\right)
+\left(  -i\gamma_{nk_{2}}^{3}\delta_{mk_{1}}\partial_{y}+m\gamma_{nk_{2}}
^{0}\delta_{mk_{1}}\right)  \right]  f_{2}^{k_{1}k_{2}},\label{sa:D_operator}
\end{equation}
and is defined in such a way as to reproduces the correct spectrum \eqref{opreg:Ham_diag_final} when acting on the ansatz \eqref{opreg:ansatz_a}, \eqref{opreg:ansatz_b} for $f_{2}^{k_{1}k_{2}}(x,y)$. The operator $\left(  \hat{h} f_{1}\right)  ^{ij}$ is fixed from the self-adjointness condition:
\begin{equation}
\langle \Phi \vert \hat{H} \Psi \rangle =\langle
\hat{H}\Phi \vert \Psi \rangle \label{sa:sa}
\end{equation}
Substituting \eqref{sa:scalar_product} into \eqref{sa:sa}, and taking into account the fact that neither the function $f_{2}^{k_{1}k_{2}}(x,y)$, nor its derivatives are continuous, after several integrations by parts one obtains the following relation:
\begin{align}\label{sa:cond1}
\alpha\left(  \frac{1}{\eta^{2}}\right)  \overset{+\infty}{\underset{-\infty
}{\int}}dv\overset{v+\eta/2}{\underset{v-\eta/2}{\int}}dx\overset{v+\eta
/2}{\underset{v-\eta/2}{\int}}dy &\left[ g_{1}^{* ij}(x,y)\left(  \hat{h}f_{1}\right)^{ij}(x,y) - \left(\hat{h}g_{1}\right)^{* ij}(x,y) f_{1}^{ij}(x,y) \right] \\
&=2i\overset{+\infty}{\underset{-\infty}{\int}}dy\left[  g_{2}^{\ast12}(x,y)
f_{2}^{12}(x,y)-g_{2}^{\ast21}(x,yf_{2}^{21}(x,y)\right]  |_{x=y+\eta}^{x=y-\eta}. \nonumber
\end{align}
Denoting:
\begin{align}
	f_{2(\pm)}^{ij}(y) & \equiv f_{2}^{ij}(y\pm\eta,y) \label{sa:f_pm} \\
	g_{2(\pm)}^{ij}(y) & \equiv g_{2}^{ij}(y\pm\eta,y), \label{sa:g_pm}
\end{align}
and using the relations $f_{2(+)}^{12}(y) = -f_{2(-)}^{21}(y)$, etc.,  the expression \eqref{sa:cond1} can be written in the form:
\begin{align}\label{sa:cond2}
2\alpha\left(  \frac{1}{\eta^{2}}\right)  \overset{+\infty}{\underset{-\infty
}{\int}}dv\overset{v+\eta/2}{\underset{v-\eta/2}{\int}}dx\overset{v+\eta
/2}{\underset{v-\eta/2}{\int}}dy &\left[ g_{1}^{* 12}(x,y)\left(  \hat{h}f_{1}\right)^{12}(x,y) - \left(\hat{h}g_{1}\right)^{\ast 12}(x,y) f_{1}^{12}(x,y) \right] \\
&=4i\overset{+\infty}{\underset{-\infty}{\int}}dy\left[  g_{2 (-)}^{\ast 12}(y)
f_{2 (-)}^{12}(y)-g_{2 (+)}^{\ast21}(y)f_{2 (+)}^{21}(y)\right]. \nonumber
\end{align}
We next fix the operator $\hat{h}$ as follows. First, we choose:
\begin{equation}
f_{1}^{12}(y,y)=[f_{2(-)}^{12}(y)+f_{2(+)}^{12}(y)]\label{sa:f1_y_y}
\end{equation}
Up to a multiplicative factor, this coincides precisely (see \eqref{opreg:ansatz_a}, \eqref{opreg:ansatz_b}) with our earlier expression
\eqref{opreg:f12_y_y}. We now define the operator $\hat{h}$:
\begin{equation}
\left(  \hat{h}f_{1}\right)^{12}(y,y)=i\left[  [f_{2(-)}^{12}(y)-f_{2(+)}
^{12}(y)\right].  \label{sa:h_op}
\end{equation}
It is easy to check that this definition respects the symmetry properties of the functions, in particular we have: $\left(  \hat{h}f_{1}\right)^{12}(y,y) = - \left(  \hat{h}f_{1}\right)^{21}(y,y)$.
Now, it is easy to see that in the limit $\eta \rightarrow 0$ the integrand in the left hand side of \eqref{sa:cond2} can be written in the form:
\begin{equation}
\left[  g_{1}^{\ast12}\left(  \hat{h}f_{1}\right)  ^{12}-\left(  \hat{h}
g_{1}\right)  ^{\ast12}f_{1}^{12}\right](y,y)  =2i\left[  g_{2(-)}^{\ast
12}f_{2(-)}^{12}-g_{2(+)}^{\ast12}f_{2(+)}^{12}\right](y)  \label{sa:final}
\end{equation}
Thus, the left and the right hand sides of \eqref{sa:cond2} coincide if we 
set $\alpha=1$ in the definition of the scalar product \eqref{sa:scalar_product}. In this way the self-adjointness of $\hat{H}$ is  shown, and, in addition, we have shown that the choice of $f_{1}^{12}(y,y)$ in \eqref{sa:f1_y_y} is in perfect agreement with our earlier analysis and with the solution we have found in \eqref{opreg:f12_y_y}. Thus, we conclude that the extra term \eqref{opreg:psi2_extra} arising in the analysis of the diagonalization of the quantum Hamiltonian $\mathcal{H}_{\Delta}$ corresponds, in the quantum-mechanical picture, to the construction of a self-adjoint extension in order to guarantee self-adjointess of the quantum-mechanical Hamiltonian $\hat{H}$.
 
\section{Conclusion}\label{sec:conclusion}

In this paper we have considered the equivalence theorem for the $AAF$-model, which asserts the invariance of the $S$-matrix under field redefinitions.  The consideration of such field redefinitions is motivated by the possibility to reduce the complicated Dirac brackets for the components of the fermion field of the original $AAF$ theory to the standard Poisson brackets. In such a way one achieves a significant simplification when considering the integrability properties of the $AAF$ model from the point of view of the inverse scattering method. We have shown here the invariance of the $S$-matrix under such transformation by appropriately modifying the perturbative proof of \cite{Bergere:1975tr}, as well as performing the direct diagonalization of the corresponding Hamiltonian. The latter has been achieved by regularizing the singular operator products at the same point by means of the Sklyanin's product. This procedure leads to a non-trivial realization of the $n$-particle states and the corresponding quantum-mechanical wave functions, which have complicated continuity properties. We show that neither the wave function nor its derivatives are continuous functions and satisfy rather complicated discontinuity equations. We were able to solve the later to show the integrability of the model and also that the $S$-matrix is unchanged.

An interesting problem to consider is whether a field transformation that would allow us to reduce a non-ultralocal theory to an ultralocal one. The $AAF$ model serves as a representative example of a classically integrable model which exhibits most of the difficulties associated with quantization of such integrable models. We have shown that the direct diagonalization requires regularization of the singular product of operators at the same point as well as the construction of the self-adjoint extensions. It is not known how to use the quantum inverse scattering method for non-ultralocal models, which classically can be described by the $(r,s)$ pair in the case of the simpler non-ultralocal models \cite{Maillet:1985ek}. In the case for the $AAF$ model the situation seems to be even more complicated, and the non-ultralocal structure is such that the classical algebra of Lax operators requires already three matrices $(r,s_{1}, s_{2})$. Interestingly, the algebra of transition matrices for both cases can still be described by a $(u,v)$ pair of matrices, constructed from the $(r,s_{1}, s_{2})$ matrices.  Since the quantization of such models is not currently known, one can try to perform  a gauge transformation of the Lax pair in order to reduce the model to an ultralocal one. This does not seem to be possible for the majority of interesting models due to the limited class of allowed transformations. On the other hand, if the proof of the equivalence theorem from the first principles can be given in some generality, then we are allowed to consider a much larger class of transformations without the necessity to perform each time the direct diagonalization.\footnote{Although we have shown here how to perform a direct diagonalization of the quantum Hamiltonian \eqref{opreg:hamiltonian}, it is clear that this is a very non-trivial and technically difficult procedure if we have to do it for every transformation.}  One could in principle, in this case, try to reduce the $AAF$ model, or any other non-ultralocal model, to an ultralocal one by considering the most general field transformation order by order. 

Another point that was not addressed in the paper is the proof of renormalizability of the $AAF$ model within the $BPHZ$ scheme. The equivalence theorem considered in \cite{Bergere:1975tr}, and which we had adopted for our model, relies on renormalizability, which we had simply assumed based on (not strictly proven) integrability and the symmetries associated with it. The fact that the quantum Hamiltonian can be diagonalized gives a strong indication that this is indeed the case. However, we have given here the proof for the false vacuum, and one has to still reconstruct the physical vacuum and the states. It would be interesting to show the renormalizability of the $AAF$ model strictly and complete the proof. For two-dimensional integrable quantum field theories, the proof of the equivalence theory must somehow be simpler. Indeed, it was shown in \cite{Parke:1980ki} that it is enough to find only two quantum conserved commuting charges in order to guarantee the factorization of the $S$-matrix. Thus, provided these two charges, one can only show the invariance of the two-particle scattering $S$-matrix. The calculation of the latter is a much simpler task, as was shown for example in \cite{Thacker:1980ei,Klose:2006dd,Melikyan:2008cy,Melikyan:2011uf}. These and related problems will be considered in future publications.



\appendix
\addcontentsline{toc}{section}{Appendices}
\section{Appendix: Invariance of Green's function} \label{sec:appendix}
\renewcommand{\thesubsection}{\Alph{subsection}}

We start by considering a set of arbitrary functions $\zeta_{1}(\psi,\bar{\psi}),\zeta_{2}(\psi,\bar{\psi}),\ldots, \zeta_{s}(\psi,\bar{\psi})$ and ${\Lambda}(\psi,\bar{\psi})$. Then, following \cite{Bergere:1975tr}, one can obtain the following equation:\footnote{For any functionals ${\Lambda}$ and ${\Gamma}$ of $\psi$, $\bar{\psi}$ and their higher derivatives, one defines:
\begin{equation}
\frac{\delta \Lambda}{\delta f_{j}}\equiv\sum_{n=0}^{\infty}\frac{(-1)^{n}}{n!}\partial_{\mu_{1}...}\partial_{\mu_{n}}\frac{\partial \Lambda}{\partial\partial_{\mu_{1}}...\partial_{\mu_{n}}f_{j}}\label{green:def1}
\end{equation}
where $j=1,2$ e $f_{1}=\psi$ and $f_{2}=\bar{\psi}$.

\begin{equation}
\frac{\vec{\delta}\Lambda}{\delta f_{j}}\Gamma\equiv\sum_{n=0}^{\infty}\frac{1}{n!}\frac{\partial \Lambda}{\partial\partial_{\mu_{1}}...\partial_{\mu_{n}}f_{j}}\partial_{\mu_{1}}...\partial_{\mu_{n}}\Gamma.\label{green:def2}
\end{equation}}
\begin{equation}
\left\langle T\left[\int dy^{2}\left\{ \frac{\vec{\delta}\mathscr{L}_{\rho}(\psi,\bar{\psi})}{\delta\psi}{\Lambda}\right\} (y){\Gamma}(x)\right]\right\rangle =i\left\langle T\left[\left\{ \frac{\vec{\delta}{\Gamma}}{\delta\psi}{\Lambda}\right\} (x)\right]\right\rangle \label{green:eom}
\end{equation}
where
\begin{align}
{\Gamma}(x)=&\prod_{i=1}^{s}\zeta_{i}(\psi,\bar{\psi})(x_{i}),\\
\left\{ \frac{\vec{\delta}{\Gamma}}{\delta\psi}{\Lambda}\right\} (x)  =&\sum_{i=1}^{s}(-1)^{i+1}\left\{ \frac{\vec{\delta}\zeta_{i}}{\delta\psi}{\Lambda}\right\} (x_{i})\prod_{j\neq i}\zeta_{j}(x_{j}).
\end{align}
Here, the factor $(-1)^{i+1}$ arises due to  anticommutativity of the fields.
Choosing the function ${\Gamma}(x)$ as follows:
\begin{equation}
{\Gamma}(x)=\prod_{i=1}^{s}\left\{ \zeta_{i} \left( \psi^{(\rho)}(\psi,\bar{\psi}),\bar{\psi}^{(\rho)}(\psi,\bar{\psi}) \right) \right\} (x_{i})\label{green:trans}
\end{equation}
where $\psi^{(\rho)}(\psi,\bar{\psi})$ and $\bar{\psi}^{(\rho)}(\psi,\bar{\psi})$ are given by \eqref{green:t1-1} e \eqref{green:t2-1} respectively, the equation analogous to \eqref{green:eom}, and written for some order $M$ of the parameter $\rho$, becomes:
\begin{align}
\left\langle T\left[\int dy^{2}\left\{ \frac{\delta\mathscr{L}(\psi^{(\rho)},\bar{\psi}^{(\rho)})}{\delta \psi^{(\rho)}}\left(1+\rho\frac{\vec{\delta}F}{\delta\psi}\right){\Lambda}+\frac{\delta\mathscr{L}(\psi^{(\rho)},\bar{\psi}^{(\rho)})}{\delta\bar{\psi}^{(\rho)}}\left(\rho\frac{\vec{\delta}\bar{F}}{\delta\psi}\right){\Lambda}\right\} (y){\Gamma}(x)\right]\right\rangle _{M}^{\mathscr{L}_{\rho}} & =\label{green:emt-1}\\
=i\left\langle T\left[\left\{ \frac{\vec{\delta}\Gamma}{\delta \psi^{(\rho)}}\left(1+\rho\frac{\vec{\delta}F}{\delta\psi}\right){\Lambda}+\frac{\vec{\delta}{\Gamma}}{\delta\bar{\psi}^{(\rho)}}\left(\rho\frac{\vec{\delta}\bar{F}}{\delta\psi}\right){\Lambda}\right\} (x)\right]\right\rangle _{M}^{\mathscr{L}_{\rho}}.\nonumber
\end{align}
Similarly, one obtains the equation corresponding to $\bar{\psi}$:
\begin{align}
\left\langle T\left[\int dy^{2}\left\{ \frac{\delta\mathscr{L}(\psi^{(\rho)},\bar{\psi}^{(\rho)})}{\delta\bar{\psi}^{(\rho)}}\left(1+\rho\frac{\vec{\delta}\bar{F}}{\delta\bar{\psi}}\right)\bar{\Lambda}+\frac{\delta\mathscr{L}(\psi^{(\rho)},\bar{\psi}^{(\rho)})}{\delta \psi^{(\rho)}}\left(\rho\frac{\vec{\delta}F}{\delta\bar{\psi}}\right)\bar{\Lambda}\right\} (y)\Gamma(x)\right]\right\rangle _{M}^{\mathscr{L}_{\rho}} & =\label{green:emt-2}\\
=i\left\langle T\left[\left\{ \frac{\vec{\delta}\Gamma}{\delta\bar{\psi}^{(\rho)}}\left(1+\rho\frac{\vec{\delta}\bar{F}}{\delta\bar{\psi}}\right)\bar{\Lambda}+\frac{\vec{\delta}\Gamma}{\delta \psi^{(\rho)}}\left(\rho\frac{\vec{\delta}F}{\delta\bar{\psi}}\right)\bar{\Lambda}\right\} (x)\right]\right\rangle _{M}^{\mathscr{L}_{\rho}}\nonumber
\end{align}
The superscript $\mathscr{L}_{\rho}$ is a reminder that the correlation functions are computed using the general Lagrangian  \eqref{green:lagrang_rho} for any parameter $\rho$.
The last two equations have a more complicated form in comparison to the case of a single bosonic field considered in \cite{Bergere:1975tr}. The main difference is the mixing of the terms $\frac{\vec{\delta}\bar{F}}{\delta\psi}$
and  $\frac{\vec{\delta}F}{\delta\bar{\psi}}$ in the above equations. We will see shortly that this mixing induces some modifications in the proof of the theorem.

The trick to show that the Green function is invariant under changes  variable \eqref{green:t1-1} and \eqref{green:t2-1}, consists of deriving the transformed Green's function  in relation to the  parameter $\rho$  and showing after some algebraic manipulations that the derivative of the Green function with respect to the parameter $\rho$ is zero, thus, showing the independence of Green's function on the parameter $\rho$. It is easy to see that:
\begin{equation}
\frac{\partial}{\partial\rho}\left\langle T\left[\Gamma(x)\right]\right\rangle _{M}^{\mathscr{L}_{\rho}}=\left\langle T\left[\int dy^{2}\frac{\partial\mathscr{L}_{\rho}}{\partial\rho}(y)\Gamma(x)\right]\right\rangle _{M-1}^{\mathscr{L}_{\rho}}+\left\langle T\left[\left\{ \frac{\vec{\delta}\Gamma}{\partial \psi^{(\rho)}}F+\frac{\vec{\delta} \Gamma}{\delta\bar{\psi}^{(\rho)}}\bar{F}\right\} (x)\right]\right\rangle _{M-1}^{\mathscr{L}_{\rho}}\label{green:derivada}
\end{equation}
The first term on the right side of the equation (\ref{green:derivada}) can be transformed as follows. Using the relation:
\begin{equation}
	\frac{\partial\mathscr{L_{\rho}}}{\partial\rho}(y)=\left\{ \frac{\delta\mathscr{L}(\psi^{(\rho)},\bar{\psi}^{(\rho)})}{\delta \psi^{(\rho)}}F+\frac{\delta\mathscr{L}(\psi^{(\rho)},\bar{\psi}^{(\rho)})}{\delta\bar{\psi}^{(\rho)}}\bar{F}\right\} (y)
\end{equation}
we can write:
\begin{equation}
	\left\langle T\left[\int dy^{2}\frac{\partial\mathscr{L}_{\rho}}{\partial\rho}(y) \Gamma(x)\right]\right\rangle _{M}^{\mathscr{L}_{\rho}}=\left\langle T\left[\int dy^{2}\left\{ \frac{\delta\mathscr{L}(\psi^{(\rho)},\bar{\psi}^{(\rho)})}{\delta \psi^{(\rho)}}F+\frac{\delta\mathscr{L}(\psi^{(\rho)},\bar{\psi}^{(\rho)})}{\delta\bar{\psi}^{(\rho)}}\bar{F}\right\} (y) \Gamma(x)\right]\right\rangle _{M}^{\mathscr{L}_{\rho}}\label{green:rel}
\end{equation}
Using equations (\ref{green:emt-1}) and  (\ref{green:emt-2}) repeatedly by  substituting $(\Lambda,\bar{\Lambda}) = (\Lambda_{i},\bar{\Lambda}_{i})$, where the sequences $(\Lambda_{i}, \bar{\Lambda}_{i}), \, i=0 \ldots n$ are given by the following iterative formulas:
\begin{equation}
  \begin{split}
	  \Lambda_{0}&=F;\\
\Lambda_{1}&=\left(-\rho\frac{\vec{\delta}F}{\delta\psi}\right)\Lambda_{0}+\left(-\rho\frac{\vec{\delta}F}{\delta\bar{\psi}}\right)\bar{\Lambda}_{0};\\ \Lambda_{2}&=\left(-\rho\frac{\vec{\delta}F}{\delta\psi}\right)\Lambda_{1}+\left(-\rho\frac{\vec{\delta}F}{\delta\bar{\psi}}\right)\bar{\Lambda}_{1};\quad \quad \nonumber\\ 
&..., \nonumber \\ \Lambda_{n}&=\left(-\rho\frac{\vec{\delta}F}{\delta\psi}\right)\Lambda_{n-1}+\left(-\rho\frac{\vec{\delta}F}{\delta\bar{\psi}}\right)\bar{\Lambda}_{n-1};\quad  \end{split}
\quad \quad
  \begin{split}
	  \bar{\Lambda}_{0} &=\bar{F},\\
\bar{\Lambda}_{1} &=\left(-\rho\frac{\vec{\delta}\bar{F}}{\delta\bar{\psi}}\right)\bar{\Lambda}_{0}+\left(-\rho\frac{\vec{\delta}\bar{F}}{\delta\psi}\right)\Lambda_{0}, \label{green:Lambda_n_sequence}\\
\bar{\Lambda}_{2}&=\left(-\rho\frac{\vec{\delta}\bar{F}}{\delta\bar{\psi}}\right)\bar{\Lambda}_{1}+\left(-\rho\frac{\vec{\delta}\bar{F}}{\delta\psi}\right)\Lambda_{1},\\
&..., \nonumber \\
\bar{\Lambda}_{n} &=\left(-\rho\frac{\vec{\delta}\bar{F}}{\delta\bar{\psi}}\right)\bar{\Lambda}_{n-1}+\left(-\rho\frac{\vec{\delta}\bar{F}}{\delta\psi}\right)\Lambda_{n-1},\nonumber
  \end{split}
\end{equation}
one can show, using \eqref{green:rel}, that the equation \eqref{green:derivada} can be written in the following form:
\begin{align}
	\left\langle T\left[\int dy^{2}\frac{\partial\mathscr{L}_{\rho}}{\partial\rho}(y)\Gamma(x)\right]\right\rangle _{M}^{\mathscr{L}_{\rho}} & =\left\langle T\left[\int dy^{2}\left\{ \frac{\delta\mathscr{L}(\psi^{(\rho)},\bar{\psi}^{(\rho)})}{\delta \psi^{(\rho)}}\Lambda_{n}\right\} (y)\Gamma(x)\right]\right\rangle _{M}^{\mathscr{L}_{\rho}}+\nonumber \\
	& +\left\langle T\left[\int dy^{2}\left\{ \frac{\delta\mathscr{L}(\psi^{(\rho)},\bar{\psi}^{(\rho)})}{\delta\bar{\psi}^{(\rho)}}\bar{\Lambda}_{n}\right\} (y)\Gamma(x)\right]\right\rangle _{M}^{\mathscr{L}_{\rho}}+\nonumber \\
	& +i\left\langle T\left[\left\{ \frac{\vec{\delta}\Gamma}{\delta \psi^{(\rho)}}\left[F+\Lambda_{n}\right]\right\} (x)\right]\right\rangle _{M}^{\mathscr{L}_{\rho}}+\nonumber \\
	& +i\left\langle T\left[\left\{ \frac{\vec{\delta}\Gamma}{\delta\bar{\psi}^{(\rho)}}\left[\bar{F}+\bar{\Lambda}_{n}\right]\right\} (x)\right]\right\rangle _{M}^{\mathscr{L}_{\rho}}\label{eq:fim}
\end{align}
Equation (\ref{eq:fim}) is true for all $n\geqslant0$,
if the last two terms on the right are equal to zero when $n=0$. 
Note that $\Lambda_{n}$ is of the order $\rho^{n}$. We will see later that this is important to show the invariance of the Green function under change of variables. The iterative construction of $\Lambda$ and $\bar{\Lambda}$ which we have used here is much more complicated than the construction used in the bosonic model \cite{Bergere:1975tr}. We can rewrite (\ref{green:derivada}) as:
\begin{align}
	\frac{\partial}{\partial\rho}\left\langle T\left[\Gamma(x)\right]\right\rangle_{M}^{\mathscr{L}_{\rho}} & =i\left\langle T\left[\int d^{2}y\left\{ \frac{\delta\mathscr{L}(\psi^{(\rho)},\bar{\psi}^{(\rho)})}{\delta \psi^{(\rho)}}\Lambda_{n}\right\} (y)\Gamma(x)\right]\right\rangle _{M-1}^{\mathscr{L}_{\rho}}+\nonumber \\
	& +i\left\langle T\left[\int d^{2}y\left\{ \frac{\delta\mathscr{L}(\psi^{(\rho)},\bar{\psi}^{(\rho)})}{\delta\bar{\psi}^{(\rho)}}\bar{\Lambda}_{n}\right\} (y)\Gamma(x)\right]\right\rangle _{M-1}^{\mathscr{L}_{\rho}}+\nonumber \\
	& +\left\langle T\left[\left\{ \frac{\vec{\delta}\Gamma}{\delta \psi^{(\rho)}}\Lambda_{n}\right\} (y)\right]\right\rangle _{M-1}^{\mathscr{L}_{\rho}}+\left\langle T\left[\left\{ \frac{\vec{\delta}\Gamma}{\delta\bar{\psi}^{(\rho)}}\bar{\Lambda}_{n}\right\} (y)\right]\right\rangle _{M-1}^{\mathscr{L}_{\rho}}\label{eq:fim-1}
\end{align}
In the equation \eqref{eq:fim-1} it is always possible to choose $n\geqslant M$, thus, the right side of the equation (\ref{eq:fim-1})  is equal to zero, 
since  $\Lambda_{n}$ and $\bar{\Lambda}_{n}$ and of the order $\rho^{n}$. Thus, the Green's function is invariant under the change of variables \eqref{green:t1-1}
and  \eqref{green:t2-1}. Then we find:
\begin{equation}
	\frac{\partial}{\partial\rho}\left\langle T\left[\prod_{i=1}^{s}\left\{ \zeta_{i} \left(\psi^{(\rho)},\bar{\psi}^{(\rho)}\right)\bar{\zeta}_{i}\left(\psi^{(\rho)},\bar{\psi}^{(\rho)} \right) \right\} (x_{i})\right]\right\rangle _{M}^{\mathscr{L}_{\rho}}=0,\label{green:zero}
\end{equation}
which shows that the matrix  $\left\langle T\left[\prod_{i=1}^{s}\left\{ \zeta_{i} \left( \psi^{(\rho)},\bar{\psi}^{(\rho)} \right)  \bar{\zeta}_{i}\left( \psi^{(\rho)},\bar{\psi}^{(\rho)} \right) \right\} (x_{i})\right)\right\rangle _{M}^{\mathscr{L}_{\rho}}$
is independent of $\rho$. Setting $\rho=0$ and  $\rho=1$ we see that, for all $M$:
\begin{equation}
	\left\langle T\left[\prod_{i=1}^{s}\left\{ \zeta_{i}(\psi,\bar{\psi})\bar{\zeta}_{i}(\psi,\bar{\psi})\right\} (x_{i})\right]\right\rangle ^{\mathscr{L}}=\left\langle T\left[\prod_{i=1}^{s}\left\{ \zeta_{i}(\psi+F,\bar{\psi}+\bar{F})\bar{\zeta}_{i}(\psi+F,\bar{\psi}+\bar{F})\right\} (x_{i})\right]\right\rangle _{M}^{\mathscr{L}_{T}}.\label{green:green_app}
\end{equation}
A key point to stress is that the original Lagrangian $\mathscr{L}$ \eqref{aafov:lag_relativistic} has a manifestly Lorentz invariant  form, while the transformed Lagrangian $\mathscr{L}_{T}$ in \eqref{green:lagrang_T} does not. Nevertheless, the relation above shows that Green's functions still exhibit Lorentz invariance, provided the functions $\zeta_i(\psi, \bar{\psi})$ are well-defined tensors.

\section*{Acknowledgments} The work of A.M. is partially supported by CAPES. The work of E.P. is supported by FAPESP grant 2009/17918-5. The work of V.O.R is supported by CNPq grant 304116/2010-6 and FAPESP grant 2012/51444-3.

\bibliographystyle{utphys} 
\bibliography{aaf_equivalence_theorem_final_v1}
\end{document}